\documentclass[preprintnumbers,amsmath,amssymb,showkeys]{revtex4}

\usepackage{graphicx,placeins}
\usepackage{dcolumn}
\usepackage{bm}
\usepackage{subfigure}
\usepackage{placeins}

\begin{document}

\title{Gene switching rate determines response to extrinsic perturbations in a transcriptional network motif}

\author{Sebastiano de Franciscis}
\affiliation{%
European Institute of Oncology, Department of Experimental Oncology\\
Via Ripamonti 435, I20141 Milano, Italy}%

\author{Giulio Caravagna}

\affiliation{%
Universit\`{a}  degli Studi di Milano-Bicocca,\\
 Dipartimento di Informatica, Sistemistica e Comunicazione, \\
 Viale Sarca 336, 20126 Milano, Italy. 
}

\author{Alberto d'Onofrio\footnote{Previous affiliation: Dept of Experimental Oncology, European Institute of Oncolog, Via Ripamonti 435, I20141 Milan (Italy)}}
 \email{alberto.donofrio@i-pri.org}
\affiliation{%
International Prevention Research Institute, \\
15 Chemin du Saquin, Ecully, 69130, France.\\
}%

\date{\today}

\begin{abstract}
It is well-known that gene activation/deactivation dynamics may be a major source of randomness in genetic networks, also in the case of large concentrations of the transcription factors. In this work, we  investigate the effect of realistic extrinsic noises acting on gene deactivation in a common network motif - the positive feedback of a transcription factor on its own synthesis - under a variety of settings, i.e., distinct cellular types, distribution of proteins and properties of the external perturbations. At variance with standard models where the perturbations are Gaussian unbounded, we focus here on bounded extrinsic noises to better mimic biological reality. Our results suggest that the gene switching velocity is a key parameter to modulate the response of the network. Simulations suggest that, if the gene switching is fast and many proteins are present, 
an irreversible noise-induced first order transition is observed as a function of the noise intensity. If noise intensity  is further increased a second order transition is also observed. When gene switching is fast, a similar scenario is observed even when few proteins are present, provided that larger cells are considered, which is mainly influenced by noise autocorrelation. On the contrary, if the gene switching is slow, no fist order transitions are observed. In the concluding remarks possible implications of the irreversible transitions for cellular differentiation are briefly discussed.
\end{abstract}
\keywords{Gene switching; extrinsic noise; intrinsic noise; bounded noises; phase transitions; stochastic bifurcations; cell differentiation; multistability; transcription factors}

\maketitle

\section{Introduction}

Gene activation and deactivation are at the base of the dynamics of transcriptional networks. However, it is only after the pioneering investigations by Ko \cite{Ko} and by Kepler and Elston \cite{KeplerElston} that their fundamental role in shaping the dynamics of such networks is starting being understood \cite{Lip2005, Kaern, Lip06, Lip13, LipP53}. In this work, we investigate the effect of {\em realistic extrinsic noises} acting on the gene deactivation mechanism in a basic network motif - the positive feedback of a transcription factor on its own synthesis - under different scenarios. More precisely, we consider the case of different cellular types hosting the network, the presence of different amounts of proteins, and various properties of the  external perturbations.

However simple (and at large extent idealised) it may be, yet this biological model summarises some of the key concepts in Systems Biology: {\em nonlinearity}, {\em  multi-stability} \cite{ZH1,BK,Iannaccone,Zhdanov2,Xiong,ZH4,ZH3,verd2014}
 (whose relevance {  to model induction phenomena and cellular differentiation} \cite{Iannaccone,Zhdanov2,Xiong,ZH4,ZH3,verd2014} was first stressed in the sixties \cite{GlassKauffman,Griffith,Simon,Thomas}),  feedback mechanisms \cite{ThomasDAri,IglesiasIngalls}, {\em stochasticity} - both in intrinsic and extrinsic form -  and the above mentioned key role of gene switching dynamics. The problem we study is also inherently {\em multi-scale}, being characterised by at least three different temporal and numerical scales: the slow/fast gene switching rates, the small/large number of copies of the transcription factor - this, for instance, is characteristic of eukaryotic cells or bacteria - and the small/large autocorrelation times of the extrinsic noise. 
 The first two scales are present also in the baseline case of isolated network, i.e., in the absence of extrinsic noise, and play an essential role in determining the dynamics of the unperturbed system, as previously stressed \cite{Lip13}. The third scale  is specifically introduced by the inclusion of  bounded coloured perturbations \cite{NoiPlos,dOnofrioBoundedNoiseBook}.

From the complex interplay of all these scales, it is not surprising that the very same biological network might behave quite differently. In other words, the transcriptional circuitry is only a part of the biological phenomenon, which emerges both from the motif itself and the particular combination of considered scales.

In order to fully appreciate the role of gene switching in modulating the dynamics of genetic networks, we provide an example taken from \cite{KeplerElston,Lip13}. Namely, it is generally thought that  a transcriptional network where a large number of proteins and mRNAs are observed can be  modelled with differential equations, in the idealised case of absent extrinsic noise. On the contrary, stochastic fluctuations - sometimes very large - can be predicted if the rate of switching of at least one of the genes involved in the network is slow, with respect to the characteristic times of the remaining part of the circuit. Thus, this phenomenon is one of the leading cause of stochasticity in genetic networks, as stressed by Elston and Kepler in \cite{KeplerElston} and by Lipniacki and colleagues \cite{Lip2005,Lip06,Lip13,LipP53}. This previously unaccounted stochasticity interplays with the nonlinearity of genetic networks and, sometimes, may induce new emergente behaviours   able to explain experimental observations \cite{KeplerElston,Lip2005,Kaern,Lip06,Lip13,LipP53}.

The above mentioned absence of extrinsic noises is a high level abstraction, even when stochasticity induced by genes switching between their ``on'' and ``off'' states  is considered (i.e., an ``internal'' phenomenon). This abstraction  is useful to establish relevant baseline behaviours, but is scarcely realistic. Indeed, no network is so isolated to always neglect its interplay with all the other intracellular  networks, as well as its interactions with signals coming from the extracellular word. Thus, we believe that a crucial step towards a better understanding of genetic networks is the inclusion of extrinsic stochastic noises, so to investigate the networks' response to various perturbations. 

However, modelling extrinsic perturbations and their effects is by no means trivial. In the past, external stochastic effects were often considered disturbances obfuscating a real signal, to be biologically controlled by those pathways working as a low-pass analog filter, as in  radiophysics \cite{detw,RaoWolfArkin}. For these reasons, a number of theoretical and experimental investigations focused on the existence of noise-reducing sub-networks \cite{detw,Thattai,Becskei1}. Thinking of noises as pure nuisances, as traditionally meant in telecommunication engineering, requires careful considerations. Indeed, in such a case a perturbed monostable network  should exhibit  fluctuations around a unique deterministic equilibrium.  However, non-equilibrium statistical and chemical physics showed that the real scenario is far more complex. 

Indeed, in the seventies Horsthemke and Lefever  seriously challenged the above mentioned mono-stability/mono-modality correspondence.  They succeeded in showing that, in  many cases, nonlinear systems monostable in absence of external  noises exhibit multimodal equilibrium  probability densities, in presence of random Gaussian disturbances \cite{hl}. They termed these  ``noise-induced-transitions'', a phenomenon shown to be  relevant also to bio-molecular networks \cite{hasty,Arkin}. In other words, the concentration of the chemicals in a network undergoing such transitions can stochastically fluctuate around more than one point in the phase space\,\footnote{It is worth noting that   earliest  works on such transitions were based on  ``white'' noises, i.e., a noise with no temporal correlation (see SM for details). White noise fluctuations are highly idealised and  not completely appropriate in our setting where, most likely,  the noise is ``coloured'', e.g., the Ornstein-Uhlenbeck noise,  and has  correlation-dependent properties that   white noises are missing \cite{hl}. Coloured noises allow observing, for instance, reversible transitions from mono- to bi-modality and other phenomena \cite{wiolindenberg}. Though more appealing, however,  coloured unbounded noises are still inadequate models of biological phenomena, since positivity and finiteness of such perturbations  can not be guaranteed (see, e.g., the critical commentaries in  \cite{dongan,dOnofrioBoundedNoiseBook}).}.

Besides extrinsic noises, experimental studies evidenced the importance of internal stochastic effects on many important biomolecular networks when transcription factors, and/or proteins and/or mRNAs  are present in a small number of copies \cite{siggia,Becskei2}. Moreover, it was shown that RNA production is not continuous, but instead it has the characteristics of {\em stochastic bursts} \cite{cfx}. Thus, a number of investigations has focused on this internal, i.e. {\em intrinsic},  form of stochasticity,   often termed ``intrinsic noise'' \cite{G77,ThattaiPNAS}. It was shown - both theoretically and experimentally - that also this form of randomness might  induce multi-modality in the probability distribution of chemicals \cite{Arkin,To}\,\footnote{Note, however, that the similarity among  intrinsically stochastic systems and  systems affected by extrinsic Gaussian noises was very well known since decades, both in statistical and chemical physics.  Indeed, this had been theoretically demonstrated by approximating the exact Chemical Master Equation with an appropriate Fokker-Planck equation \cite{Gardiner,Hanggi,Hanggi2,G80}, an approach leading to the Chemical Langevin Equation \cite{Gillespie00}.}.

The potential of exploiting {  internal and} external noises  to fluctuate around  points that are non-equilibrium when noise is absent{  , as well as to switch from an equilibrium to another, are not} just a mathematical curiosity, as {  they} can  be  biologically   functional  (see, e.g., \cite{eldar,losick,bucetaplosone, GarciaOjalvo2012, Tsimring} and references therein). For example, this might allow a  network to reach biochemical {  ``equilibrium''} configurations otherwise unaccessible \cite{Arkin,eldar,losick}. Phenotype variability in cellular populations is, for example, probably the most important macroscopic effect of intracellular noise-induced {  phenomena} \cite{eldar,NoiPlos}.

Some body of research has been devoted to investigations concerning the interplay between intrinsic  and extrinsic forms of stochasticity.  From a modelling perspective Swain and coworkers  stressed many important effects, although without observing emergent multi-modal phenomena, in \cite{msb08}. The above study is  remarkable since: $(i)$ it has outlined the role of  noise autocorrelation, $(ii)$ it has stressed what anticipated above, namely that     unbounded  noises may induce artefacts with no sense in a biological setting, e.g., negative kinetic parameters\footnote{To avoid this, the
authors employed a log-normal positive noise which, however,  allows   the perturbed parameter to attain excessively large values, being only bounded to the ``left". It is worth noting that, from the data analysis point of view, You and collaborators \cite{You} and Hilfinger and Paulsson \cite{hilfpaul} recently proposed interesting methodologies to deconvolve the contributions of extrinsic noise in some nonlinear networks.}.

Thus, extrinsic  bounded noises seem and adequate tool to model extrinsic perturbations, in biochemical settings. Motivated by some biological examples, intrinsically stochastic  models of nonlinear networks where combined with such noises in \cite{GiulioDon12PONE12,NoiNatComp}. In that case,  the chemical reaction rates were perturbed by coloured bounded extrinsic noises modelled by  biochemical state-dependent systems of stochastic differential equations\footnote{In the same paper, a suitable extension of  the popular Gillespie algorithm was defined \cite{G77},  taking into the account the co-presence of the two different kinds of stochasticity. This new modelling approach allowed to investigate network motifs of interest in cellular biochemistry (e.g., futile cycles and toggle switches), eventually showing potentially functional noise-induced-transitions. Similarly,  the correctness of the quasi-steady state approximation for  classical  Michaelis-Menten enzymatic reactions was investigated. However, the relevance of bounded noises goes far beyond cellular biochemistry, and got recent attention in various disciplines (see, e.g., \cite{dOnofrioBoundedNoiseBook} and references therein).}. Concerning the parameters to model  noise, besides  autocorrelation, both  amplitude and stationary probability density of the noise are of the utmost importance. Indeed, the literature has shown that various stochasticity-induced phenomena may vary as a function of these noise features  \cite{dongan,dOnofrioBoundedNoiseBook} and, thus, without experimental information on the stochastic fluctuations for the problem in study, a theoretical work must deal with various models of noise.

The analysis of the behaviour of our simple biological multi-scale model is by no means trivial. As we shall see in the next, from the combination of the various scales characterising the network, various modelling approaches will be adopted, leading to different scenarios never studied before. In fact,  the simplest case of large number of proteins and  absence of extrinsic influences, when deterministic  network models can be adopted, was already studied by  Griffith \cite{jsgriffith} and later by Thomas and D'Ari \cite{ThomasDAri},  by means of boolean networks. In both cases the hypothesis of absence of a baseline activation rate was assumed. More recently, this constraint has been relaxed by Smolen, Baxter and Byrne \cite{sbb}. {Their model was employed to investigate the differentiation of WB15-M cells in response to BMP2 stimulation by Wang {\em et al.} in the theoretical/experimental paper \cite{Iannaccone}. Quite interestingly, bimodal behaviour of cellular differentiation obtained in the experimental part of \cite{Iannaccone} were reproduced by means of impulsive random changes of the transcription factor level.  In a follow-up, Smolen, Baxter and Byrne } considered the role of intrinsic noise and  other important transport-induced phenomena in this transcriptional circuit \cite{sbb99}. {Analytical investigations of the effects of {\it intrinsic} stochasticity were recently investigated by employing of the classical white-noise-based chemical Langevin equations and its innovative coloured-noise-based version \cite{bucetaplosone,frigolaplosone}. This approach allowed researchers to  propose important inferences on the effect of such a noise in the process of cellular differentiation \cite{bucetaplosone}, and on the dynamics of lactose \cite{frigolaplosone}.} In the framework of continuous approximation of protein concentration, the effects of extrinsic white noises on protein   production  have been studied by \cite{PreSBB} and  \cite{PhysaSBB}, in a biological setting partially unclear. For the sake of comparison, in SM we provide a detailed commentary of these two papers.
To conclude, the interplay between intrinsic and extrinsic unbounded noise affecting protein production  in  a self-transcription network with sharp and smooth positive feedback has been considered by Assaf {\em et al.} \cite{AssafPRL}. In there, the  biological setting (i.e., unbounded and state-dependent perturbation of protein production when the genes are ``on'' in absence of gene-switching) is substantially different from ours (i.e., bounded perturbation of the gene-deactivation dynamics), as we comment in SM.  

The paper is organised as follows: in \S \ref{sec:model} we introduce the network motif and our modelling approach according to the considered scales. Then, in \S \ref{sec:network}, we describe and comment simulations. Concluding remarks, including some future perspective derived by our numerical analysis, close this work. In supplementary materials we provide the necessary background on the mathematical definition of bounded noises and analytical calculations used to derive the the parameters settings used thorough the analysis.

\section{A transcriptional network with positive feedback}\label{sec:model}

\begin{figure}[t]
\centerline{
\includegraphics[width=0.8\textwidth]{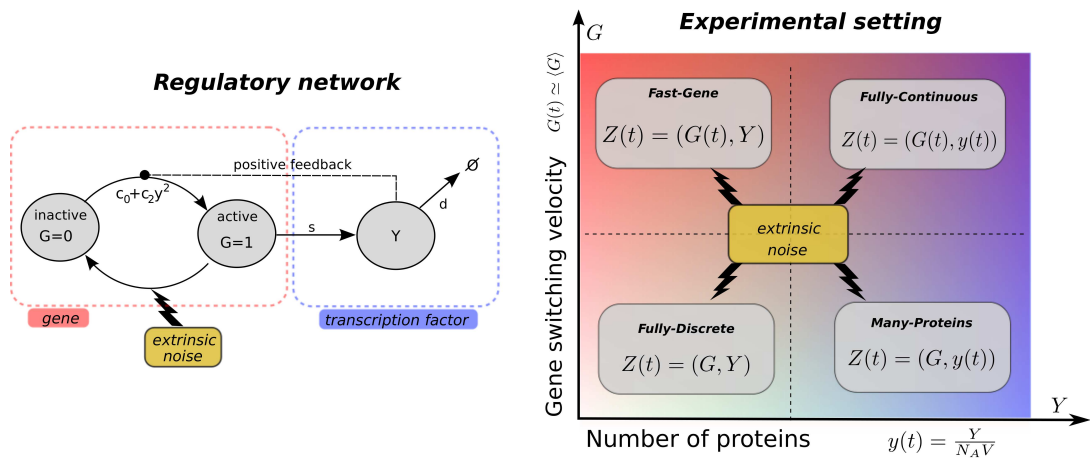}
}
\caption{\textbf{The transcriptional network and the modelling setting.} Left: the model consists of a gene switching from active/inactive states, with a transcription factor acting with positive feedback on the activation. An extrinsic noise abstractly models  the possible unknown interactions of the network with its environment. For instance, if noise has an oscillatory effect on the deactivation rate, its role is to model  a certain - potentially unknown - protein synthesised in an oscillatory regime, which competes with the network protein  making its deactivation rate oscillate. Right: we model the network under four different settings, according to the gene on/off switching velocity and the number protein involved. This requires to use different mathematical representation of the network, whose  robustness is investigated under the effect of bounded extrinsic noises.}
	\label{fig:network}
\end{figure}

We consider a simple transcriptional network constituting  of a gene $G$ and its protein product, a transcription factor with positive feedback on its own gene. Specifically, here we assume that the deactivation rate of the gene is under the influence of  bounded  extrinsic perturbations (Figure \ref{fig:network}, left). 

The protein  numbers will be denoted as $Y$, and its concentration as $y=Y/N_{A}V$, where $V$ is the cell volume and $N_{A}$ Avogadro's number. Usually, one assumes  the transcription factor to be produced by $n=2$ copies of $G$. However, in pathological cases more/less copies of $G$ might be needed: e.g., due to gene deletions  it may be $n=1$ \cite{Stranger}, while in tumour cells it may be $n>2$, due to genetic amplification \cite{Cappuzzo}. In the applications, we set $n=2$.

In the above network a single gene copy could be {\em active}, i.e. producing its transcript with rate $s$, or {\em silent}, so at each time instant  we model the gene as a binary variable $G\in\left\{0,1\right\}$. We assume that the transcript protein positively feedbacks on its own production, thus proteins do not act in the process of gene deactivation. 

We denote as $c(y) = c_0 + c_2y^{2}$ and $b_0$ the, respectively, activation and deactivation rates of the gene, while the protein degradation rate is $d$ (Figure \ref{fig:network}, left). Thus the positive feedback is modelled with an additive term in activation rate $c_2y^{2}$, and $b_0$ and $c_0$ are the baseline rates of deactivation and activation.
Extrinsic perturbations affect $b_0$, yielding a time-varying deactivation rate:
\begin{equation}
b_0(t)=b_\ast [1+ \xi(t)]\, ,
\end{equation} 
where $b_\ast$ is the average value of the deactivation rate and $\xi(t)$ is a {\em bounded noise} such that $b_0(t)>0$ and $\langle b_0(t) \rangle = b_\ast$. Every noise is characterized by an amplitude, $B \in [0,1]$, and an autocorrelation time, $\tau > 0$; see the supplementary materials (SM) for a detailed description of the two types of bounded noises that we consider in this work. Roughly, $B$ and $\tau$ define the noise intensity and  its speed in changing value. From a modelling perspective, we consider noise as an abstract representation of the possible unknown interactions of the network with its environment. For instance, when $\xi(t)$ has an oscillatory effect on $b_\ast$  we are abstractly modelling a certain protein synthesized in an oscillatory regime, which competes with the protein $Y$ making its deactivation rates oscillate. In principle, though noises can affect all the network components, we restrict our analysis to the effects on deactivation, since an exhaustive characterization of the noise effects is out of the scope of the present work. 

As mentioned in the introduction, models of a network similar to the one we consider here were previously studied. For instance, in \cite{jsgriffith,sbb99} the case of deterministic modelling was considered, while in \cite{sbb99,Lip13}  intrinsic stochastic effects were included. In addiction, the effects of unbounded white/coloured Gaussian  noises modelling biological mechanisms not clearly defined were investigated in \cite{PreSBB,PhysaSBB,AssafPRL} (see detailed comments in the SM). The model of noise used in those studies - i.e., unbounded - is nowadays considered inappropriate in the context of biological systems, see, e.g.,\cite{msb08,NoiPlos,NoiNatComp}. Bounded noises, instead, which  seem to be much more suitable, are used hereby to model a clear  biological  phenomenon. In this paper we consider the network of figure \ref{fig:network}  in four different experimental settings emerging from two, out of the three possible,  scales mentioned in the introduction, namely: $(i)$ {\em the slow, respectively fast, velocity of gene on-off switching} and $(ii)$ {\em the small, respectively large,  number of  proteins involved}. Then, in each of  those settings we will assess the network response to extrinsic bounded noise.
This makes the network a simple - but powerful -  {\em multi-scale} system  where both temporal and numerical scales coexist. To the best of our knowledge, a combination of different experimental settings and bounded extrinsic noises acting on the gene deactivation mechanism was not studied before.

Mathematics provide us  a framework for {\em exact modelling} of the above network, which is based on the idea of counting explicitly the number of proteins and states of the genes in the network \cite{G77}. Such a framework is, by its constituting ingredients, unaware of the underlying physical hypothesis on the involved number of proteins. More precisely, it assumes that numerical and temporal scales are somehow similar, a fact which is hard to guarantee in  a general setting as the one studied here. Unfortunately, this leads to known  {\em scalability issues} which emerge when that framework is used, e.g., to model a multi-scale system as the one above. Pragmatically,  an exhaustive analysis of an exact model might turn out to be impractical, both at the level of the  state-space, which might be too large to sample many initial conditions, or even more dramatically at the level of the single network simulation. However, the different settings allow us to use approximated mathematical representations of the network, sometimes dropping precision for complexity of the model analysis. On one side, this drawback is anyway unavoidable, as suggested by the lot of efforts in developing efficient techniques for model analysis. On the other side, this is methodologically correct, being  the result of applying careful mathematical considerations on the physics of the network itself.

Our study, which includes four distinct modelling scenarios, is in line with the recent important work by Jaruszewicz {\em et al.} \cite{Lip13}, with the noteworthy exception that in that paper a noise-free setting and a single-gene network were considered. We remark that including noises does not allow much analytical investigation of the network, as instead was possible in \cite{Lip13}. The modelling techniques that we employ in the next sections are,  in order of decreasing complexity:
\begin{itemize}
\item {\em Slow gene switching and small number of copies of the protein}, which requires us to {\em account explicitly for all the copies of genes/proteins} via a pure Continuous-Time Markov Chain (CTMC) representation of the network - augmented with bounded noises - as defined in \cite{GiulioDon12PONE12}. We present it in \S \ref{sec:model} model A. This is the exact representation of the network  and, as anticipated, it can  be  efficiently analyzed only when a few hundreds of proteins are present and the time-scales of the involved events are similar;

\item {\em Fast gene switching velocity and small number of protein copies}, which allows us to make a {\em steady-state hypothesis on the gene dynamics}. This model is mathematically similar to  model A (with slow gene switching and small number of copies of the protein), but counting exactly solely the proteins. We present it in \S \ref{sec:model} model B. This  first approximation of the exact model is necessary since, if we were here to model explicitly the gene-switching process, most of the simulation time would be spent in modelling gene's activation/deactivation, downgrading the overall performance;

\item {\em Slow gene switching velocity and large number of protein copies}, which leads to a {\em steady-state hypothesis on the protein concentration}. In this model the gene state $G(t)$ is a stochastic process (as in model A), whereas the dynamics of the proteins is approximated via a differential equation driven by $G(t)$. In other words the model (which is hybrid because involves a stochastic process and a differential equation) is a classical Piecewise-Deterministic Markov Process \cite{Davis}, augmented with noise. We present it in \S \ref{sec:model} Model C. In this case,  possible deadlocks emerging when simulating thousands of proteins in a CTMC representation are avoided;

\item {\em Fast gene switching velocity and large number of protein copies}, which allows us to model the network with coupled differential equations, plus extrinsic noise, constituting the model presented in \S \ref{sec:model} model D. In this case, if we were to use the CTMC representation, deadlocks would surely  emerge in simulating the model, making its analysis unfeasible.

\end{itemize}

In general, these approaches result is a series of models of decreasing complexity both in terms of  mathematical representation and  cost of simulation. In the following, we describe each of the models outlined above.

\FloatBarrier

\newcommand{\ZZ}{\mathbf{Z}}
\newcommand{\zz}{\mathbf{z}}

\subsection{\bf  Model A: Slow gene switching and small number of proteins} 

Here we study the network when the switching rate of the genes is slow and the number of molecules of the transcription factor is small. Let the state of the system be $\ZZ(t)=(G,Y)$. Each state $\ZZ(t)=\zz$ is characterized by a jump rate defined by the following {\em time-inhomogeneous birth-death process}.

The events of the model are described in Table \ref{table:modelA}, where we used the notation $G \rightarrow G+1$ to denote that the event increase by one the number of $G$ in the time interval $(t,t+dt)$, i.e. an ``off'' gene becomes ``on''. Similarly, $Y \rightarrow Y-1$ models a protein degradation with rate $d$.

Here $G$ and $Y$ determine the discrete state of the embedded Markov Chain, and the overall process is time-inhomogeneous  because of the effect of noise on the  {\tt (deactiv.)} event, i.e., via $b_0(t)$
 which contains the  noise term $\xi(t)$. Because of this the jump equation to generate a  simulation is to be solved numerically, as discussed in \cite{GiulioDon12PONE12,dOnofrioBoundedNoiseBook}. If $\xi(t)=0$, standard algorithms could have been used to simulate the embedded CTMC, e.g. \cite{G76}, which would have been time-homogeneous since $b_0(t)=b_\ast$. Note that, since the gene deactivation rate must be strictly positive, in this and in the following three models it is impossible to represent the extrinsic noise by means of a Gaussian (or non-Gaussian) unbounded noise.

\begin{table}[!ht]\center
\caption{\bf {Model A:} Slow gene switching and small number of proteins.}
\begin{tabular}{  r c l |  r c l}
  \hline
  \multicolumn{3}{c}{\em Gene  (events)} &   \multicolumn{3}{c}{\em Protein  (events)} \\
  \hline
 & {\em effect} & \;{\em rate equation}  & & {\em effect} & \;{\em rate equation}  \\
 {\tt (deactiv.)} & \;$G \rightarrow G-1$ & \;$a_1(\zz,t)=b_0(t) G$ & {\tt (transcr.)} & \;$Y \rightarrow Y+1$ & \;$a_4(\zz,t)= s N_{A}V G$  \\  
 {\tt (activ.)} & \;$G \rightarrow G+1$ & \;$a_2(\zz, \cdot)=c_0[n-G]$ & {\tt (degr.)}& \;$Y \rightarrow Y-1$ &\;$a_5(\zz,t)= d Y$  \\
 {\tt (feed.)}& \;$G \rightarrow G+1$ & \;$a_3(\zz,\cdot)= \dfrac{c_2 Y^2 [n-G]}{(N_{A}V)^2 }$ \\
  \hline
\end{tabular}
\begin{flushleft}
\end{flushleft}
\label{table:modelA}
\end{table}

\FloatBarrier

\subsection{\bf  Model B: Fast gene switching and few proteins} 

We consider the case of a switching rate of the gene to be fast enough to satisfy  
\begin{align*}
c_0\gg d, &&b_0(t)\gg d .
\end{align*}
Since  in this case gene switching  is very quick, we can  assume  that $G(t)\approx \langle G(t)\rangle$ \cite{KeplerElston,Lip13}, thus 
\begin{equation}\label{GeneEqHyb2}
\langle G(t) \rangle = \dfrac{n \left(c_0 +\dfrac{c_2 Y^2(t)}{N_{A}^2V^2}\right)}{b_0(t)+c_0 +\dfrac{c_2 Y^2(t)}{N_{A}^2V^2}} = 
 \dfrac{n \left[c_0 +c_2 y^2(t)\right]}{b_0(t)+c_0 +c_2 y^2(t)}\, .
\end{equation}
Here we are switching to a model state  $\ZZ(t)=(G(t),Y)$ which contains one continuous component, the genes, and aggregate the effects of the  {\tt (deactiv.)}, {\tt (activ.)}  and  {\tt (feed.)} events in a unique mean-field approximation of $\langle G(t) \rangle$. Thus, in this case, the network itself is hybrid (i.e., joint discrete-continuous), regardless of the noise effects, and it is described by the processes in Table \ref{table:modelB}.

In this case, the technique introduced in \cite{GiulioDon12PONE12} is extended to account for $\langle G(t) \rangle$ as a mean-field variable, along the lines of usual stochastic hybrid systems. 

\begin{table}[!ht] \center
\caption{ {\bf Model B:} Fast gene switching and few proteins.}
\begin{tabular}{  r | c l l }
  \hline
{\em Gene  (mean-field)} &   \multicolumn{3}{c}{\em Protein  (events)} \\ \hline
$\langle G(t) \rangle = \dfrac{n \left[c_0 +c_2 y^2\right]}{b_0(t)+c_0 +c_2 y^2}$ & {\tt (transcr.)} & \;$Y \rightarrow Y+1$ & \;$a_4(\zz,t)= s N_{A}V \langle G(t) \rangle$ \\
 &  {\tt (degr.)}& \;$Y \rightarrow Y-1$ &\;$a_5(\zz,\cdot)= d Y$  \\
  \hline
\end{tabular}
\begin{flushleft}
\end{flushleft}
\label{table:modelB}
\end{table}

\FloatBarrier

\subsection{\bf  Model C: Slow gene switching and large number of proteins} 

When the switching velocity of the gene dynamics is low but the number of molecules is large, one can replace the protein rate equations  with the following  mean-field model for the protein density $y(t)$
\begin{equation}\label{rde}
\dot{y}= s G - d y\, ,
\end{equation}
and consider a model state  $\ZZ(t)=(G,y(t))$. As a consequence, in terms of jumps we are aggregating the effects of the  {\tt (transcr.)}  and  {\tt (degr.)}  in  $y(t)$, and both  {\tt (deactiv.)} and   {\tt (feed.)} events are time-inhomogeneous. Such a model is described by the processes in Table \ref{table:modelC}.

In this case, which is symmetrical to the previous one, the network itself is a hybrid discrete-continuous model because of the mean-field approximation $y(t)$. As before, the technique introduced in \cite{GiulioDon12PONE12} must be extended to account for $y(t)$, as in the previous case.

\begin{table}[!ht]\center
\caption{ {\bf Model C:} Slow gene switching and large number of proteins.}
\begin{tabular}{  r  c l | c }
  \hline
\multicolumn{3}{c}{\em Gene  (events)} &   {\em Protein  (mean-field)} \\ \hline
  {\tt (deactiv.)} & \;$G \rightarrow G-1$ & \;$a_1(\zz,t)=b_0(t) G$ & \\
   {\tt (activ.)} & \;$G \rightarrow G+1$ & \;$a_2(\zz, \cdot)=c_0[n-G]$ & $\dot{y}= s G - d y$\\
  {\tt (feed.)}& \;$G \rightarrow G+1$ & \;$a_3(\zz,t)= c_2 y^2(t)[n-G]$ &  \\
  \hline
\end{tabular}
\begin{flushleft}
\end{flushleft}
\label{table:modelC}
\end{table}

\FloatBarrier

\subsection{\bf  Model D: Fast switching velocity and many proteins} 

In the case where the switching velocity of the gene dynamics is fast and the number of molecules is large, then one can substitute to the embedded CTMC the two approximations presented in the previous sections. This yields the mean-field model described in Table \ref{table:modelD}.

Notice that in this case this equation is coupled with the equation modelling noise, thus the overall model is still a stochastic process.
The equation for protein density which we have in this case is 

\begin{equation}\label{mysde} 
y^{\prime}= s \frac{n (c_0+c_2 y^{2})}{(c_0+c_2 y^{2})+b_0(1+\xi(t))}  - d y,
\end{equation}
where $\xi(t)$ is a bounded noise.

Note that in the baseline case of absence of extrinsic noise ($\xi(t)=0$) in literature the resulting ordinary differential equation is often written \cite{sbb,PreSBB,PhysaSBB,AssafPRL} in the {algebraically equivalent} form
\begin{equation}\label{sbb}
\dot y= R_{b} + \frac{K_f y^2}{K_d + y^2}  - d y,
\end{equation}
which is known as the Smolen-Baxter-Byrne model \cite{sbb}, whose parameters are defined as follows 
\begin{align*}
n s = R_{b}  + K_f, && b_0 = \dfrac{c_0 K_f}{R_{b}}, &&  c_2 = c_0 \dfrac{R_{b}  + K_f }{K_d R_{bas}}\, .
\end{align*}
{It is important to note that the parameters of equation (\ref{sbb}) cannot be dealt with as if they were independent. This important aspect is discussed in  SM. }

\begin{table}[!ht]\center
\caption{ {\bf Model D:} Fast switching velocity and many proteins.}
\begin{tabular}{  c | c  }
  \hline
{\em Gene  (mean-field)} & {\em Protein  (mean-field)} \\
  \hline
$\langle G(t) \rangle = \dfrac{n \left[c_0 +c_2 y^2(t)\right]}{b_0(t)+c_0 +c_2 y^2(t)}$ &
 $\dot{y}= s \langle G(t) \rangle - d y$   \\
  \hline
\end{tabular}
\begin{flushleft}
\end{flushleft}
\label{table:modelD}
\end{table}

\FloatBarrier

\newcommand{\Probab}{{\mathcal P}}

\section{Results}\label{sec:network}

We sut up the model by adopting the values taken from the Smolen-Baxter-Byrne model \cite{sbb} $d=1 \; min^{-1} $,$R_{b} =0.4 \; min^{-1}$, $K_f = 6 \; min^{-1}$ and $K_d = 10 \; nM^{2}$ one obtains: $s=3.2  \; min^{-1}$, $c_2 = 1.6 c_0 nM^{-2}$ and $b_0 = 15 c_0 $. In absence of noise, i.e. $b_0(t)=b_\ast$, so model D is in its multi-stability region.

Note that based on the data given in \cite{sbb} one cannot identify all the original parameters. Thus, we start our investigation in the biological setting of fast gene switching and large number of protein copies (in \S \ref{sec:model} model D), which, in absence and - to some extent - in presence  of extrinsic bounded perturbations can be analytically studied (see SM). We set $c_0 = 10 d$ so that $b_0 = 150$. For this value of $b_0$ the system is multistable and has two stable equilibria at $y_L =0.6268 $, $y_H = 4.28 $ and one unstable equilibrium at $y_U =1.489 $. By assuming $b_0$ as a bifurcation variable, one gets a classical hysteresis bifurcation parameter shown in figure \ref{FullyDetDia} $(a)$-$(b)$, where one can see that the system is bistable for $b_0 \in (b_l,b_r)$ where $b_l \approx 140.5$ and $b_r \approx 174.5 $.

As remarked in the introduction, three main factors may influence the biological response to extrinsic noise, all of which we will be scrutinised in the forthcoming simulations. These are: $(i)$ the noise model, which here we restrict to the sine-Wiener and the Cai-Lin  cases (see SM and, e.g., \cite{wioII, caiwu,bobryk,dimentberg}); $(ii)$ the noise amplitude $B$; $(iii)$ the noise autocorrelation time $\tau$, which is also relevant for unbounded noises (see the  Ornestein-Uhlenbeck noise in \cite{AssafPRL, RosenfeldScience}).

In the following we will illustrate some numerical simulations of the  stationary {\em probability density function} of the number (or of the  concentration) of proteins, denoted as $\Probab_{st}(Y)$, as well as of  derived summary statistics such as the average stationary value of $Y$, $\langle Y \rangle$, and of its standard deviation, $\sigma=\sqrt{\langle Y^2\rangle-\langle Y\rangle^{2}}$. Of  course, in numerical simulations one can only measure an heuristic probability $\Probab(Y)$  at a large time $T$ which must be far larger than the characteristic times of the network in study, in order to result $\Probab(Y)\simeq\Probab_{st}(Y)$. Of  course, $T$ must also be smaller than the average lifespan of the host organism. Throughout our study  we set $T=10^4 \; min \approx 7 \; days$.

Thus the adherence of the measured heuristic density $\Probab(\cdot)$ to $\Probab_{st}(\cdot)$ ultimately depends on the velocity of convergence of the density to its stationary value.

\subsection{Model D: Fast gene switching and many proteins} \label{sec:ode model-results}

In  SM we developed an analytical study of the transient probability density of the number of proteins at time $t$, denoted $\Probab(y,t)$.

The key result of this analysis is that if $y(0)$, i.e., the initial number of proteins available at time $0$,  belongs to the basin of attraction of the low equilibrium then for small (but not infinitesimal) values of $B$, which is the noise intensity, $y(t)$ remains small, for intermediate values of $B$ $y(t)$ can undergo a jump towards the large values and this jump is \textit{irreversible} (if $B$ remains constant). Finally, if $B$ is sufficiently large, then $y(t)$ can stochastically oscillate from low to large values and vice versa. Similar scenario is predicted if $y(0)$ is large.

This mathematical analysis suggests that, assessing the influence of the noise amplitude on the stationary average $\langle y\rangle$ value of $y$, one should observe - for increasing $B$ -  before a fist-order phase  transition \cite{peliti2011statistical,SoleBook}, i.e. a sudden increase  of the average value $ \langle Y \rangle$ accompanied, in the transition point, by a sudden widening of the standard deviation $\sigma$, and then a second order phase transition, i.e. a smooth decrease of $\left<y\right>$ accompanied, in the transition point, by a widening of the observed $\sigma$.

\begin{figure}[t]
\centerline{
\includegraphics[width=0.8\textwidth]{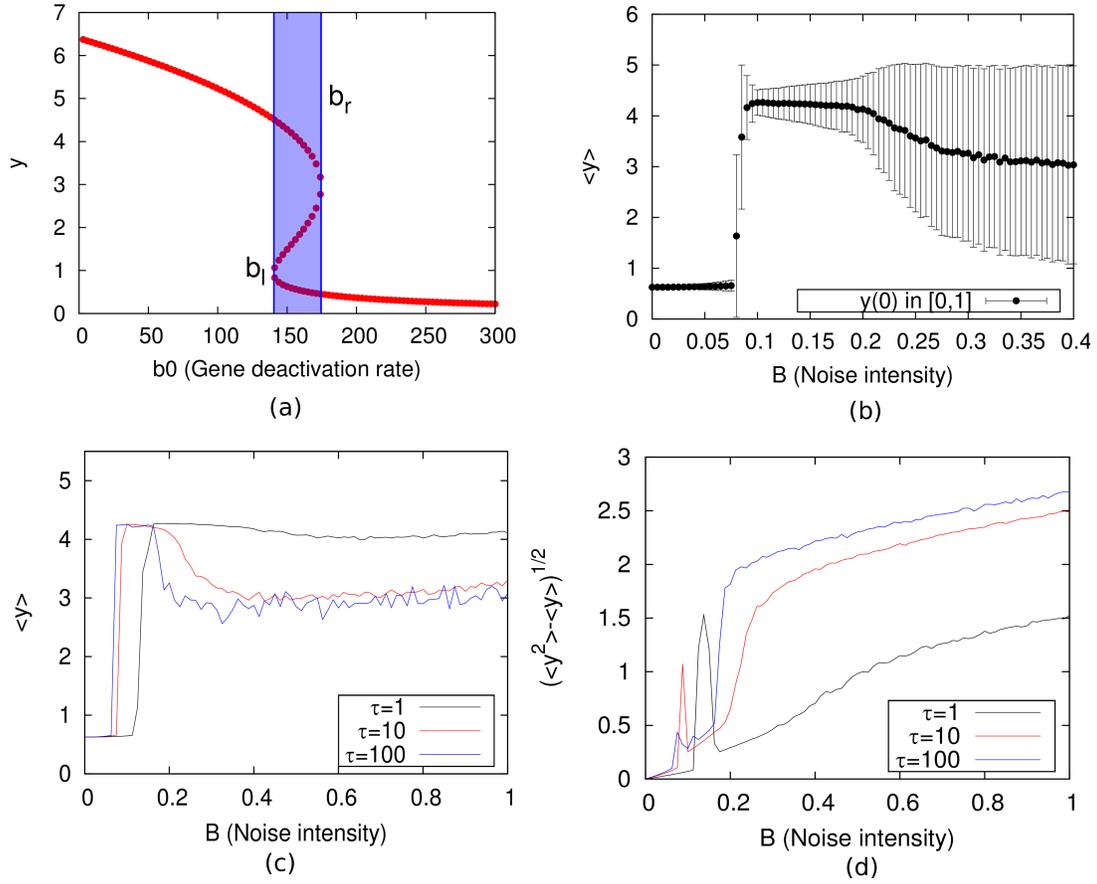}
}
\caption{{\bf  Phase diagrams when the  gene switching is fast and many proteins are present.} Panel (a): Bifurcation diagram of the protein equilibrium density against the gene deactivation rate $b_0$ for the unperturbed, fully continuous, \S \ref{sec:model} model D, with  $s=3.2  \; min^{-1}$, $c_2 = 16 \; nM^{-2}$ and $c_0 = 10 $. Note that the system is bistable for $b_0 \in (b_l,b_r)$ where $b_l \approx 140.5$ and $b_r \approx 174.5 $. Panel (b): Sine-Wiener noise perturbation with $\tau=10$. The points corresponds to simulations with initial protein density in the low protein level basin of attraction ($y(0)\in\left[0,1\right]$). The first order transition between low and high number of protein is obtained for $B\approx0.066$, corresponding to $b_0(t)\approx b_l$, the lower bound of hysteresis curve. Panel (c)-(d): we show how different autocorrelation times affect the average number of proteins by using a sine-Wiener noise with $\tau=1,10,100$. 
In both diagrams we set $c_0=10$.
}
\label{FullyDetDia}
\end{figure}

Numerical simulations confirm the theoretical predictions: a first order transition, between high and low protein densities, at $B_c \approx 0.066$ and a second order one, between high protein density and an oscillating behaviour at $B_d \approx 0.166$, as shown in Figure \ref{FullyDetDia} $(b)$ we compare  the statistical summaries (order parameters) $\langle y \rangle$ and $\sqrt{\langle y^2\rangle-\langle y\rangle^2}$ against  $B$ and $\tau$, for both sine-Wiener and Cai-Lin noises. If $B<B_c$ the equilibrium distribution is unimodal and peaked on the high or low protein level, depending on the initial number of proteins, i.e. the basins of attraction for high and low protein level are separated by the unstable fixed point, see panel $(a)$. The first order transition is also well characterised by the divergent behaviour of $\sqrt{\langle y^2\rangle-\langle y\rangle^2}$ around the transition point, see panel $(d)$. 
As shown in panels  $(c,d)$ of Figure \ref{FullyDetDia}, the increase of the  autocorrelation time $\tau$ deeply impacts on the  phase  transition regions, which get smaller while the  variance of $y$ significantly increases. In other words, the second  order transition is characterised by stochastic oscillations whose  amplitude increases with $\tau$ (at least in the range $[1,100]$  employed in our simulations). 

\begin{figure}[t]
\centerline{
\includegraphics[width=0.8\textwidth]{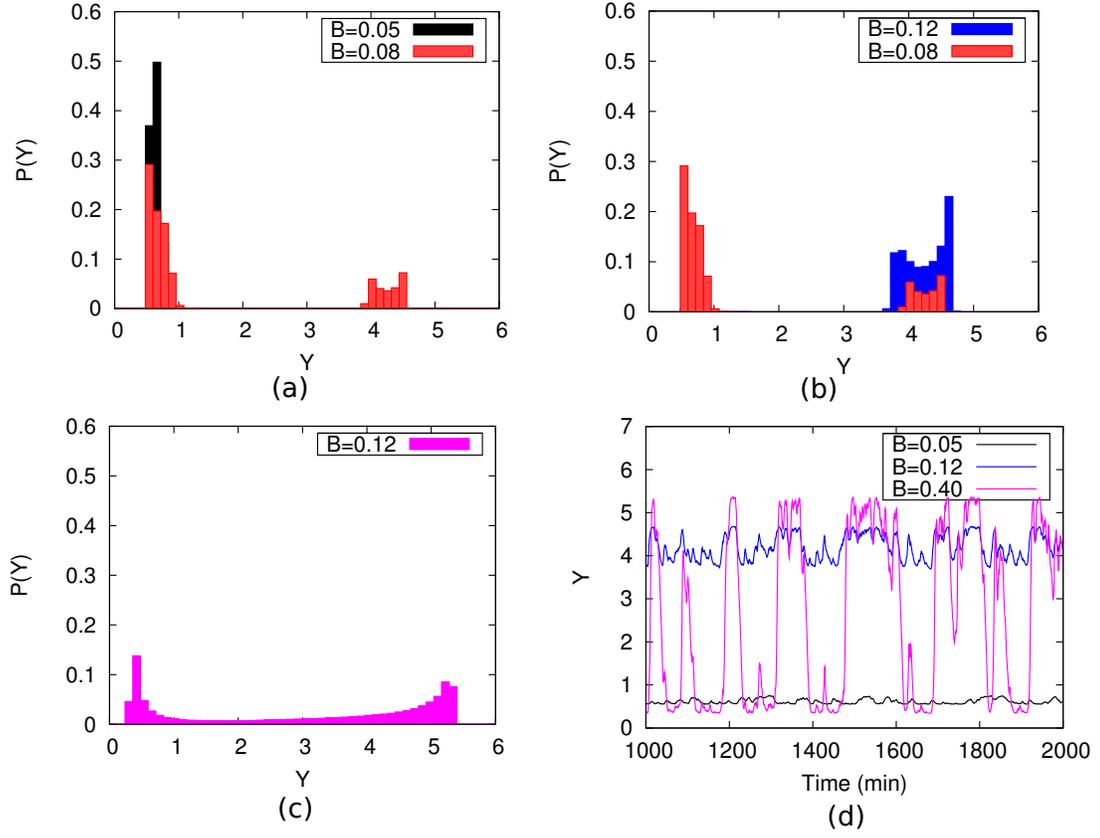}
}
\caption{{\bf  Protein distribution when the  gene switching is fast and many proteins are present.} Protein probability densities $\Probab(\cdot)$ (panels (a)-(c)) and the corresponding time series (panel (d)) for the model D,  \S \ref{sec:model}, adopting sine-Wiener noise perturbation with $\tau=10$ and various amplitudes. A first order transition between low and up protein levels emerges at $B\approx0.08$, while a second order one, between up level and oscillating up/low levels emerges for higher values, see diagram in Figure \ref{FullyDetDia} $(b)$. In all figures we set  $c_0=10$ and the protein initial number $y(0)\in\left[0,1\right]$.
}
\label{FullyDetSeries}
\end{figure}

Quite interestingly, in our biological setting the type of noise considered, i.e., the way in which the unknown competing proteins affect protein $y$, does not appear to remarkably influence such transitions (data not shown). The effect of the noise amplitude on the stationary probability of $y$ (and its time series) is shown in Figure \ref{FullyDetSeries}. For small amplitude of $B$ the fluctuations are around small values of $y$. For intermediate values the probability (and the time series, of course) jumps towards large values of $y$. Finally for even larger values of $B$ a bimodal density is observed (corresponding, in the time series, to oscillations between large and small values).

\FloatBarrier

\subsection{Model A: Slow gene switching and small number of proteins} \label{sec:markovmodel-results}

\begin{figure}[t]
\centerline{
\includegraphics[width=0.8\textwidth]{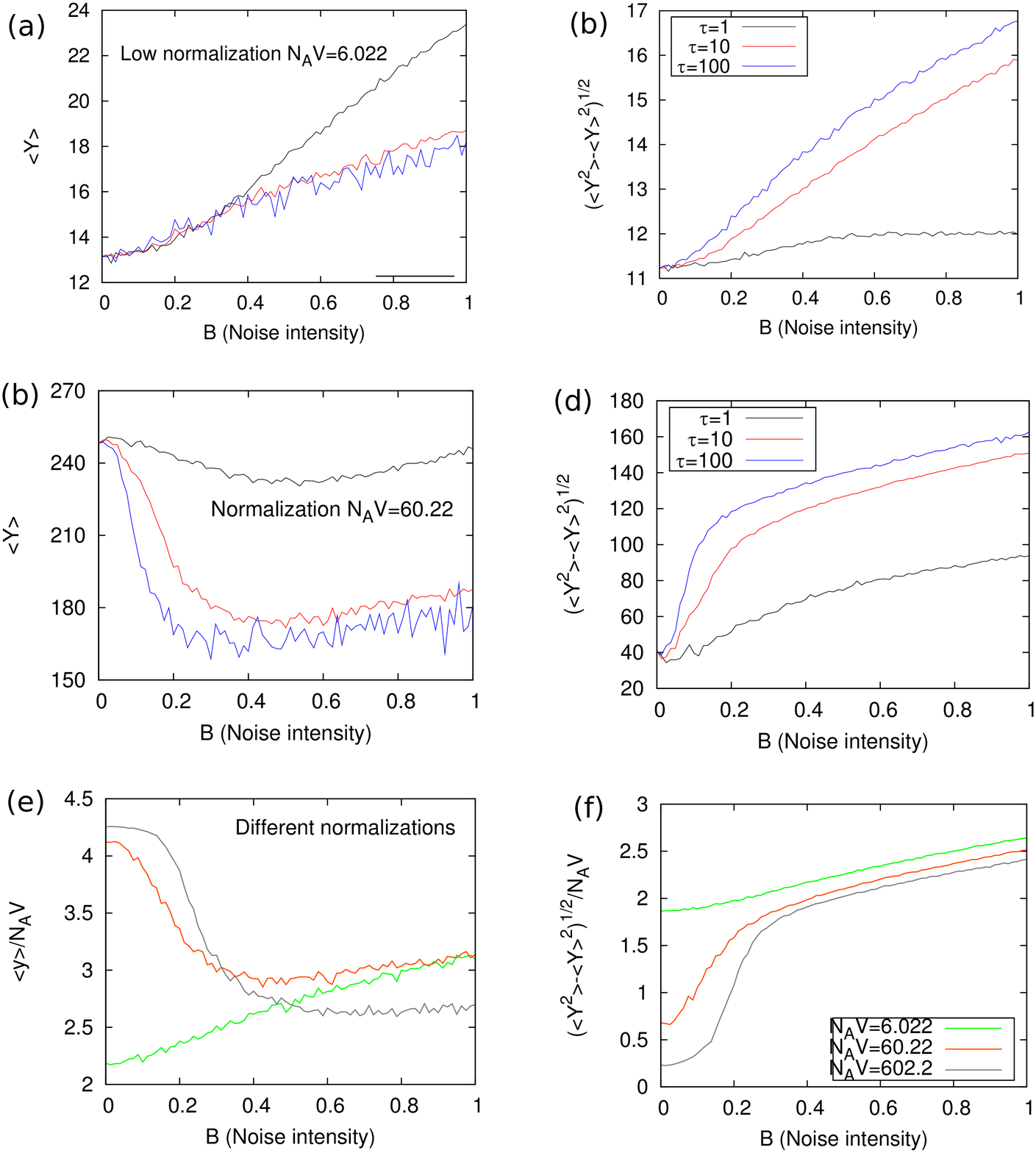}
}
\caption{{\bf  Phase diagrams when the  gene switching is slow and few proteins are present.}  A. Panels (a)-(b): average number of proteins and standard deviation of protein density, adopting a Cai-Lin noise ($z=-0.5$) with various autocorrelation values, low normalisation $N_{A}V=6.022$ and initial condition $Y(0)=1000$. Low normalisation corresponds to small the number of proteins involved in the dynamic. The typical state is oscillating between high/low protein levels, and for low $\tau$ the protein equilibrium states become ``fuzzy``, up to a point that it is not possible to distinguish among them (see also Figure \ref{FSDistriI}). Panels (c)-(d): same as panels (a)-(b) with $N_{A}V=60.22$ and initial condition $Y(0)\in\left[0,100\right]$. Increasing normalisation the second order transition between high protein level and oscillating state emerges. Panels (e)-(f): comparison of transition behaviour for different protein number normalisation, adopting a sine-Wiener noise (initial condition $Y(0)\in\left[0,10\times N_{A}V/6.022\right]$). In all diagrams we set $c_0=10$.
}
\label{FSDia}
\end{figure}

\begin{figure}[t]
\centerline{
\includegraphics[width=0.8\textwidth]{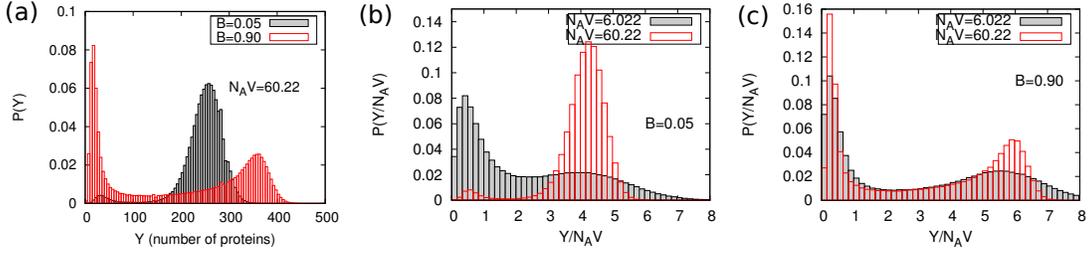}
}
\caption{{\bf  Protein distribution with slow gene switching, few proteins and sine-Wiener noise.} Panel (a):
protein density distributions, adopting sine-Wiener noise ($\tau=10$) with low normalisation $N_{A}V=6.022$. We observe the usual second order transition form high protein level to an oscillating state. Even with a very small bound value the system can rarely switch off the large protein level, resulting in a small left residual peak in the distribution. Panels (c)-(d): normalised protein density distributions $\Probab(Y/N_AV)$ for different values of normalisation $N_{A}V$. Noise amplitude $B$ enhances the peak corresponding to the low protein level and increase the up/low protein states gap. In all plots we set as protein initial condition $Y(0)\in\left[0,10\right]$.
}
\label{FSDistri}
\end{figure}

In this case, simulations of the model have been carried out by the exact algorithm proposed in \cite{GiulioDon12PONE12}. However, here we calculate the phase diagrams with an approximated algorithm, in which it is assumed that the time-scales of noise and gene-protein dynamics are very well separated. Finally, in order to check the consistency of the approximation, the probability density $\Probab(Y,t)$ is calculated with the exact algorithm. 
We compare the protein dynamic of the network with the one obtained with fast switching and many proteins, referring to the same bifurcation diagram of Figure \ref{FullyDetDia} $(b)$. In addition, in this model a normalization constant for protein density -- that determines the order of magnitude of protein number involved -- has to be set. By considering the volumes of bacteria\cite{Alberts} and eukaryotic nucleus \cite{mcgraw} in a range  of $V=10:10^3 \; \mu m^{3}$, we explore a range of values  $N_{A}V\simeq 6.022\cdot\left[1:10^2\right] \; nM^{-1}$.
In this case the first order transition predicted by  model D disappear, and for low values of $B$ the equilibrium corresponds to large or small protein level, depending from the normalisation adopted, see Figure \ref{FSDia}, panels $(a)$, $(c)$ and $(e)$. For large $N_{A}V$ we observe the usual second order transition form high protein level to an oscillating state, see Figure \ref{FSDia} panels $(c,d,e,f)$. The protein density distribution and the time series, depicted in Figure \ref{FSDistri}, reveal that, in this case, with a very small bound value ($B=0.05$) the system can rarely switch off the large protein level. In the density distributions this results in a small residual peak correspondent to low protein numbers, see Figure \ref{FSDistri} $(a)$.

For $N_{A}V = 60$ we observe an increase of the average value  $ \langle Y \rangle$. For $N_{A}V = 60$ and for $N_{A}V = 600$ a second order  transition from  a high protein level (with relatively small ratio  $\sigma/ \langle Y \rangle$) to an oscillating state (with far larger $\sigma/ \langle Y \rangle$ but a smaller $ \langle Y \rangle$),  see Figure \ref{FSDia} panels $(c,d,e,f)$. 
Thus,  the larger is number of protein involved in the system, determined by normalization, the more separate and well distinguishable are the up/low protein states. This confirms that it is crucial to use an exact modelling approach in this setting.
The  protein density distribution, depicted in Figure \ref{FSDistri} panels $(a,b,c)$, illustrate the above results. Indeed, for  very small bound value ($B=0.05$) the system rarely switch off the  large protein level. In the density distributions this results in a  ``quasi unimodal'' distribution, where  only a small residual peak at low  protein numbers is observed.

\begin{figure}[t]
\centerline{
\includegraphics[width=0.8\textwidth]{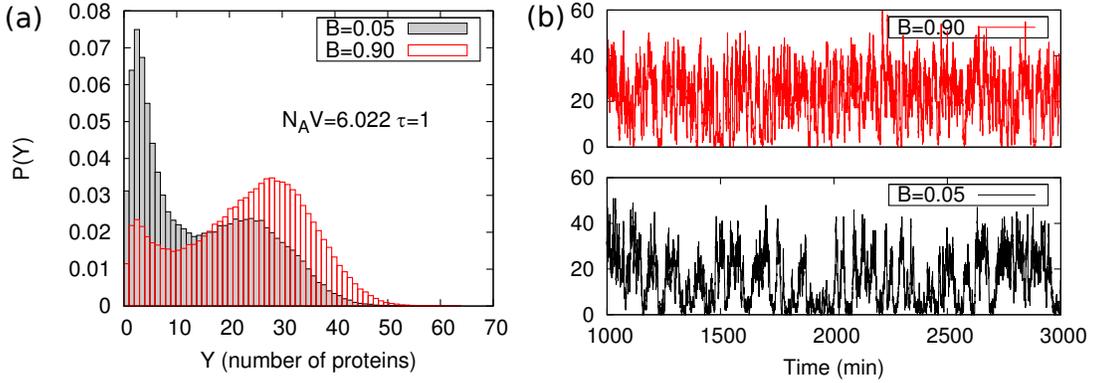}
}
\caption{{\bf  Protein distributions and corresponding time series with slow gene switching, few proteins and slow Sine-Wiener noise.} We set here $\tau=1$ for the noise and $N_{A}V=6.022$ for the normalisation. The protein initial condition is $Y(0)\in\left[0,6.022\right]$. When the noise autocorrelation time is low, $B$ has the effect to enhance the peak corresponding to the high protein level. Moreover with low normalisation, it is possible to appreciate the intrinsic stochasticity in protein numbers, that mix down the protein states, up to a point that it is not possible to distinguish among them.
}
\label{FSDistriI}
\end{figure}

As far as the effect of the autocorrelation time $\tau$ on the probability density is concerned, for $\tau\le10$ noise amplitude $B$ enhances the peak corresponding to the low protein level and increase the up/low protein states gap, see  Figure \ref{FSDistri} $(b,c)$. On the contrary, for low noise autocorrelation (see Figure \ref{FSDia} panels $(a,b)$ with $\tau=1$), $B$ has the effect to enhance the peak corresponding to the high protein level, see Figure \ref{FSDistriI}. 

\FloatBarrier

\subsection{Model B: Fast gene switching velocity and few proteins} \label{sec:pdmp1model-results}
                                                                                                                                                                                                                                                                                                                                                                                                                             
\begin{figure}[t]
\centerline{
\includegraphics[width=0.8\textwidth]{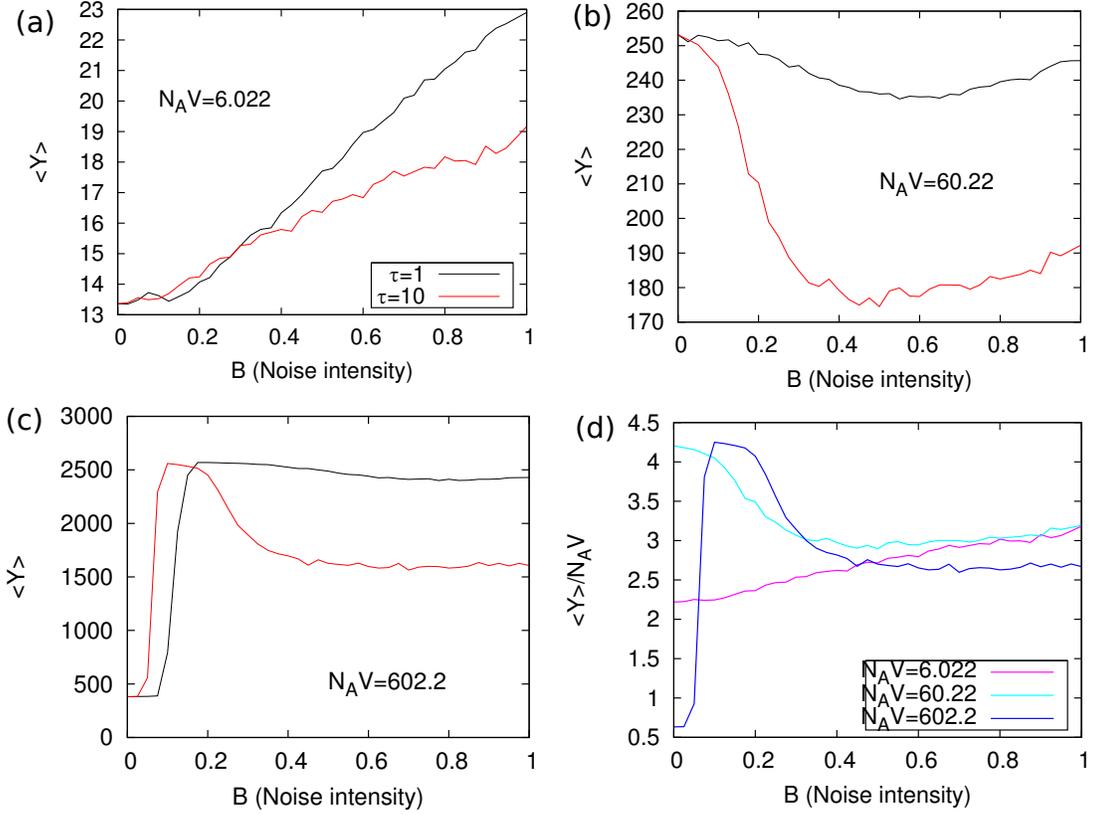}
}
\caption{{\bf  Phase diagrams when the  switching velocity is fast and few proteins are present.}  Panels (a)-(c): we use a sine-Wiener noise with different autocorrelation values and set $N_{A}V=6.022, 60.22, 602.2$, respectively. Low noise autocorrelation time $\tau$ has a double effect: it increases the amplitude of oscillating state, and enlarge the low protein region in the parameters space. Panel $(d)$: we use the same noise with $\tau=10$, and show the normalised distribution of proteins $\Probab(Y/N_AV)$ in order to compare distribution with different normalisation. When noise amplitude is small, a re-entrant transition in the normalised mean number of proteins emerges. In all panels we set $Y(0)= \langle G(0)\rangle=0$.
}
	\label{HybGenDia}
\end{figure}

\begin{figure}[t]
\centerline{
\includegraphics[width=0.8\textwidth]{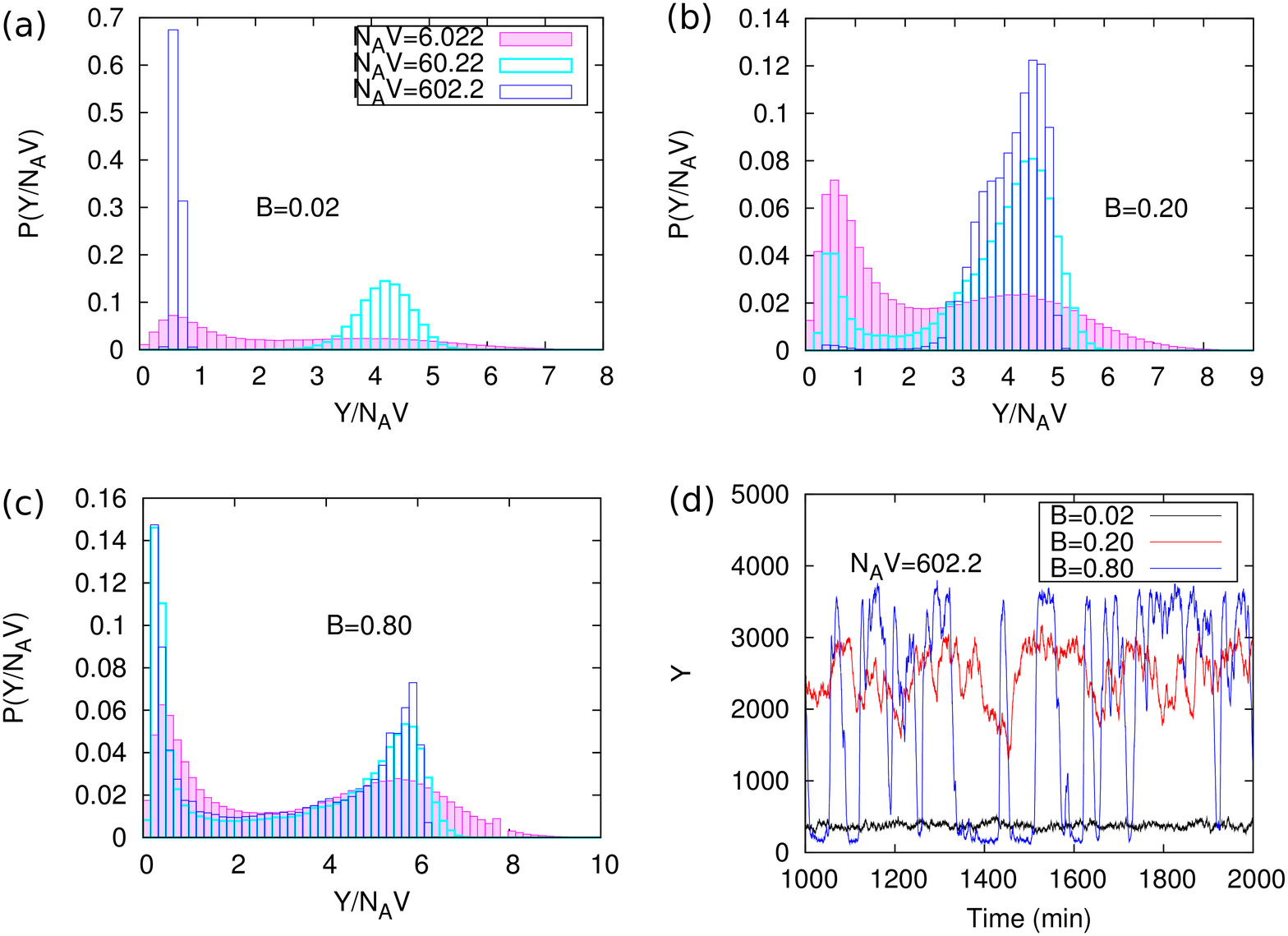}
}
\caption{{\bf  Protein distribution with fast switching velocity and few proteins.}  Panels $(a)$-$(c)$: distributions evaluated with a sine-Wiener noise and different protein normalisation, with noise amplitude $B=0.02,0.20,0.80$. For $B=0.02$ there is a re-entrant transition in normalised protein number $Y/N_AV$, since protein density distribution $\Probab(Y/N_AV)$ switches from bimodal/oscillating to unimodal/high level, and finally to unimodal/low level. Panel $(d)$:  time series generated with the same noise and  $N_{A}V=602.2$. In all panels we set $\tau=10$ and the initial conditions are $Y(0)=\langle G(0) \rangle=0$.
}
	\label{HybGenDistri}
\end{figure}

In this setting the model behaviour strongly depends on the parameter $N_A V$ and on the autocorrelation time $\tau$. Indeed, adopting noise amplitude size $B$ as order parameter, there are three different regimes (see Figures \ref{HybGenDia} and \ref{HybGenDistri}): $(i)$ for small $N_A V \approx 6$ no transition are observed, and the system is always in an oscillating state, $(ii)$ for intermediate $N_A V \approx 60$ a second order transition form large to an increasingly oscillating behavior is observed; $(iii)$ for sufficiently large $N_A V \approx 600$ we recover, as in the case of fast gene switching and large number of proteins (\S \ref{sec:ode model-results}), the same first order transition between low/high protein levels followed by the second order transition from high protein level to oscillating state. 

When noise amplitude is small, a re-entrant transition in normalised protein number $Y/N_AV$ emerges, and the normalised protein density distribution $\Probab(Y/N_AV)$ switches from bimodal/oscillating to unimodal/high level, and finally to unimodal/low level, see Figures \ref{HybGenDia} $(d)$ and \ref{HybGenDistri} $(a)$. Important consequences on cell biochemical equilibrium could be deduced from this phenomenology, in particular regarding cell mitosis, when the cell volume increase and the disaggregation of cell nucleus and the final division in two daughter cells change dynamically the normalisation term. We reserve a deep analysis of the nature of transition in $V_AN$ in future works (see conclusions in \S \ref{conclu}).
Analogously to the model A, lowering noise autocorrelation time $\tau$ enhances the low protein region in the parameters space as well as increases the amplitude of oscillating state, as depicted in Figure \ref{HybGenDia} $(b,c)$. Thus the gene switching  velocity is a factor that deeply influence the behaviour of the  system.

\FloatBarrier

\subsection{Model C: Slow gene switching velocity and many proteins} \label{sec:pdmp2model-results}

When the switching times of the genes expressing the self-regulating transcription factor are of the same or lower order than the degradation time of the protein (\S \ref{sec:model} model C), the effects of the bounded noise are very similar to the ones resulting when the switching velocity is low and few proteins are present, in the limit of high $N_{A}V$ (model A). Thus the first order transition predicted when the switching is fast and many proteins are present (\S \ref{sec:ode model-results}) here disappears, i.e. for low $B$ values the equilibrium corresponds always to high protein level. 

\begin{figure}[t]
\centerline{
\includegraphics[width=0.8\textwidth]{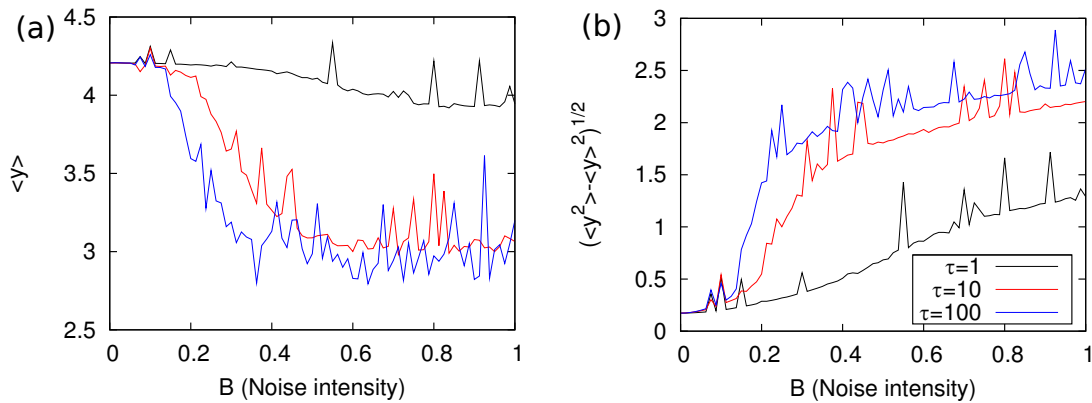}
}
\caption{{\bf  Phase diagrams with slow switching velocity and many proteins.} We use here  a Cai-Lin noise with $z=+0.5$. Analogously to the second order transition, predicted predicted by the fully discrete model A for high normalisation values, we observe a transition from an high protein level equilibrium, with a unimodal distribution, to an oscillating state, characterised by a bimodal protein distribution (see Figures \ref{FSDia} $(c,d,e,f)$ and \ref{FSDistri}). We set $c_0=10$ and initial conditions $y(0)\in\left[0,1\right]$.
}
	\label{HybDia}
\end{figure}

Analogously to the second order transition, predicted by  model A, we observe a transition from an unimodal (high protein level equilibrium) to a bimodal protein distribution with both sine-Wiener and Cai-Lin perturbations, in the range of $B\approx[0.1,0.3]$, see Figure \ref{HybDia}. This transition has the same dependence from the type of noise and its autocorrelation time as the one observed for the continuous model (compare Figures \ref{HybDia} and \ref{FullyDetDia}).

\section{Discussion and Conclusions}\label{conclu}

We performed an exhaustive computational analysis of a minimal transcriptional network (i.e., a motif), under the effect of 
realistic stochastic perturbations on the gene deactivation rate and in different experimental settings, i.e., low/high number of proteins and multiple cellular volumes. 

Results  suggest that, in general, the  gene switching velocity is the key parameter to modulate the response of the network to such perturbations. Furthermore, in the case of small number of proteins, another parameter which predicts different responses is cellular volume, which suggest that the very same network might exhibit different quantitative behaviours according to the considered cell type.

Concerning  extrinsic noises, quite surprisingly simulations suggest that little can be imputed to the particular stationary distribution  of the noise itself, e.g., horned versus bell-like. Instead, noise amplitude - which models how strong is the effect of the perturbation on the gene deactivation rate - induces a cascade of phase transitions: from first to second order, at least when gene switching is fast and a very large number of proteins is present. The same behaviour is observed even when few  proteins are present provided  cellular volume is large.

When gene switching is fast, and many proteins are present,  the first order transition becomes irreversible, i.e. once the protein concentration switches from low  to high values, there is no backward switch  unless noise amplitude is further increased. Conversely, when the gene switching is slow, no fist-order transitions were observed.

The autocorrelation time characteristic of the extrinsic noise was studied, mimicking the presence of inhibiting proteins competing for the same transcription sites of the gene, and synthesised at different velocities. Apparently, autocorrelation  affects solely  the second order (smooth) transitions by amplifying the probability to observe few proteins.

{  The bounded-noise-induced {\it irreversible} first order transitions among low and large protein levels of the transcription factor could be cautiously read as a mechanism, employed by bi-potent cells where the factor is abundant and the gene switching is very fast, for the choice of a {\it permanent} cellular fate. Indeed, for these cells the fate choice induced by the presence of intrinsic noise is missing, whereas, on the contrary, unbounded noises would in any case induce stochastic oscillations of the protein concentration (taken apart, for the sake of the discussion, the issues of biological realism). Heterogeneity of fates in a population of such cells might thus be originated by considering the amplitude of the noise  as a ``static'' random variable. Of course, these inferences are quite speculative and they need further theoretical and experimental investigations.}

{  More in general}, it is important to stress that the model here proposed is quite abstract and {  generic}. In particular, a more realistic description of the gene-switching process  might lead to biological predictions of interest. For instance,  our model neglects some important macroscopic features: mRNAs are not considered,  spatial effects \cite{Tsimring} are missing  - which were instead stressed as important {  for the motif in study since the seminal paper} \cite{sbb99} - and other issues. 

As far as future investigations are concerned, we remind that we focused on a region of parameter values that induces bistability in the deterministic case, in order to outline the possible peculiar effects of extrinsic noise. Of some interest would be, instead, a more canonical investigation of the parametric regions inducing monostability {  in the deterministic model} . Such an analysis would allow to verify whether extrinsic bounded  noises acting on the gene deactivation rate induce bimodality in the protein distribution, {  and the role of the noise autocorrelation} times.  

All of the above considerations depict a complex scenario, and motivate at least  {three}, more substantial, issues worth of further investigations. The first concerns the behaviour of the network in the context of cycling cells: indeed, since in the case of few proteins cellular volume plays a relevant role, it is natural to ask what might happen in cycling cells when this parameter varies in time, due to mitosis. The second issue is related to the dichotomy between physiological and abnormal cells. Here we investigated normal cells where the number of gene is conserved through time, i.e., $n=2$. Abnormal cells, which might be constituted of more/less copies of the gene, might allow to observe phase transitions which are unobservable when $n=2$. Along this line, first order transition might be observed, if $n$ is sufficiently large (i.e., gene amplification), and a mean field behaviour of gene switching could emerge, also when this process is slow. A positive answer to this question might lead to further investigations on the correlation of this phenomena with  proteins which are over-expressed in cancer cells.  {Finally, experimental evidences showed that the ``on-off'' two-states gene switching is often an oversimplification \cite{Blake2006,Suter2011,gutierrez2012}. Exploring multi-state and more complex mechanisms of gene dynamics is thus an important further issue. The introduction of such mechanisms might lead to an additional variability, similar to those we hypothesise for gene amplification. }

\appendix

\section{Supplementary Material}

\subsection{Critical commentary of the literature on the self-activation transcriptional network motif}

In this section we comment three previous models of the self-activation transcriptional network motif that investigated the role of extrinsic Gaussian noises, either alone \cite{PhysaSBB,PreSBB} or in combination with intrinsic fluctuations (but without gene-switching noise) \cite{AssafPRL}.

As we mentioned in the main text,  model D in absence of extrinsic noise reads as follows:
$$ 
y^{\prime}= s \frac{n (c_0+c_2 y^{2})}{(c_0+c_2 y^{2})+b_0}  - d y,
$$
{but, in the literature, e.g., \cite{sbb,PreSBB,PhysaSBB,AssafPRL}, it is often written  in the algebraically equivalent form}
\begin{equation}\label{sbbappendix}
\dot y= R_{b} + \frac{K_f y^2}{K_d + y^2}  - d y,
\end{equation}
{where $R_b$ and the sum $K_f + R_b$ can be, respectively, legitimately read as a baseline and an ``asymptotic'' production rates. However, since it holds that}
$$ R_b = n s \frac{c_0}{c_0+b_0} $$
$$ K_f = n s $$
$$ K_d = \frac{c_0+b_0}{c_2} $$
{it follows that the parameters $R_b$, $K_f$ and $K_d$ must not be dealt with as they were independent. In particular, a fluctuation of the baseline rate $R_b$ cannot be deconvolved by fluctuations in the other two parameters. }

{Unfortunately this is what happens in literature. For example,} in \cite{PhysaSBB} the Smolen-Baxter-Byrne model was studied in the framework of the above mentioned continuous approximation with fast gene-switching and large number of proteins. {The authors of \cite{PhysaSBB} analytically investigated the consequences of white stochastic oscillations affecting the baseline production rate $K_b$ and/or the parameter $K_f$. Thus, the model in \cite{PhysaSBB} reads as follows}
\begin{equation}\label{sbbphysa}
\dot y= R_{b} (1+\xi_0(t) )+ K_f (1+\xi_1(t) )  \frac{ y^2}{K_d + y^2}  - d y\, .
\end{equation}

{Unfortunately, according to  what we above showed, the stochastic differential equation (\ref{sbbphysa}) does not correspond to a biologically meaningful scenario, unless both $ \xi_0(t) = \xi_1(t)$ and, of course, the noises are bounded. We note here that, in such a particular case, eq. \cite{PhysaSBB} can be read as a model of fluctuations in the parameter $s$. }

In \cite{PreSBB} it is investigated a model where both the baseline protein production rate and the degradation rate were perturbed by white noises, yielding the following stochastic differential equation
\begin{equation}\label{SBBpre}
\dot y= R_{b} (1+\xi_0(t) )+ K_f   \frac{ y^2}{K_d + y^2}  - d (1+\xi_d(t) ) y\, ,
\end{equation}
Again, isolated oscillations of the parameter $R_b$ are not meaningful.

As briefly mentioned in the introduction, by means of semi-analytical methods, joining the Wentzel-Kramers-Brillouin (WKB) approximation and  numerical simulations, Assaf and colleagues  recently investigated the interplay between extrinsic and intrinsic noise in the circuit of a self-transcription factor with a sharp positive feedback \cite{AssafPRL}. The biological differences between their model and ours are worth to be described in some detail. 

Indeed, in their main model the extrinsic noise perturbs the production rate of the transcription factor, and the perturbation is state-dependent being  active only if the state of the protein is ``high''. 
Indeed, under the implicit hypothesis that the gene switching velocity is large they assume the following probability law for the production of the transcription factor in the time interval $(t,t+dt)$
\begin{equation}\label{Psharp}
 \Probab(Y \to Y+1) = dt \; A (\alpha_0 + (1-\alpha_0+\xi(t)) \theta(Y - Y_*) )
\end{equation}
where $\theta(.)$ is the Heavyside function and $ \alpha_0 \ll 1$. Thus if $Y(t) $ is under the threshold then the production rate of the transcription factor is unperturbed. 

The mechanism through which this state-dependent fluctuation of the production rate can be enacted is not specified. In absence of such specification,  the asymmetry of the perturbation acting on the baseline and on the large protein synthesis rates  remains unclear. 

Note that in \cite{AssafPRL} it is also briefly investigated a model where a smooth feedback is enacted (as in the Smolen-Baxter-Byrne model \cite{sbb,sbb99}):
\begin{equation}\label{Psmooth}
 \Probab(Y\to Y+1) = dt \; A \left(\alpha_0 + (1-\alpha_0+\xi(t)) \frac{Y^2}{K_d +Y^2} \right) 
\end{equation}
Thus, the perturbation adopted in \cite{AssafPRL} for such a smooth case is equivalent to the perturbation of the parameter $K_f$ alone investigated in the paper \cite{PhysaSBB} in the case $\xi_0(t)=0$. 

We note here that this equivalence can be extended to the 'sharp' model defined by \ref{Psharp} because in absence of extrinsic noise eq \ref{Psharp} is the limit case of a generalization of our model B (fast gene switching and small number of proteins) with constant $b_0$. Indeed, in the hypothesis that gene switching is fast and that the activation is caused by 'H-meric' forms (and not dimeric), proceeding as in our main text one gets (in absence of noise on $b_0$):
\begin{equation}\label{Psharpmollified0} \Probab(Y \to Y+1) = dt s N_A V \langle G(t) \rangle = dt s N_A V\dfrac{n \left[c_0 +c_2 y^H\right]}{b_0+c_0 +c_2 y^H},\end{equation}
which may be rewritten as follows:
\begin{equation}\label{Psharpmollified} \Probab(Y \to Y+1) =  dt A ( \alpha_0 + (1-\alpha_0)  \frac{(Y/Y^*)^H}{1 +(Y/Y^*)^H},\end{equation}
where
$$ Y_* = ( \frac{b_0 +c_0}{c_2})^{1/H} $$
$$ A \alpha_0 \approx s N_A V n \frac{c_0}{b_0 +c_0}. $$
Thus, if $K$ is sufficiently large \ref{Psharpmollified} reads:
$$\Probab(Y \to Y+1) \approx dt \; A \left(\alpha_0 + (1-\alpha_0) \theta(Y - Y_*) \right) $$
Summarizing, we may say that for both the sharp and the smooth models of \cite{AssafPRL} the intrinsic noise acts in a biologically unrealistic way, equal (for $K=2$) or remindful (for $K>>1$) of that investigated in \cite{PhysaSBB} in the case $\xi_0(t)=0$.

As far as the extrinsic noise is concerned, Assaf and coworkers considered an unbounded Orenstein-Uhlenbeck noise defined as follows
$$ \dot \xi  = - \frac{1}{\tau}\xi +\sqrt{ \frac{2 \sigma_{ex}}{\tau} }\eta(t) \, , $$
where $\eta(t)$ is a unitary white noise.
In this way the stationary probability density function of the noise $\xi$ is Gaussian with variance $\sigma_{ex}$.
However, from the above equations it follows that it must be 
$$ \xi(t) > 1- \alpha_0\, . $$
One might roughly consider tolerable the error if
$$ 2 \sigma_{ex} < 1 - \alpha_0\, , $$
which, however, does not seem the case in  \cite{AssafPRL}. Indeed in the simulations presented in   \cite{AssafPRL}  the employed ``stochastic bifurcation parameter'' is the ratio between the standard deviation of the Orenstein-Uhlenbeck noise and a parameter named $\mu$. This parameter is defined as the average of the (transitory) ``quasi-stationary distribution  about the high state'' \cite{AssafPRL}. Since in their simulations the high state is large (it ranges from $300$ to $5000$), even $\mu$ has to be large. Indeed, we simulated the sharp feedback model by means of the (exact) Gillespie algorithm and we got that the average value of the probability density function is very close to the high equilibrium state. This means that the standard deviation $\sigma_{ex}$ in most of the simulations reported in \cite{AssafPRL} largely exceeds $0.5 (1-\alpha_0) $.

\subsection{Background on temporal bounded noises and autocorrelation times}\label{noises}

Temporal bounded noises can be generated either via {\em stochastic differential equations}, e.g. \cite{wioII, caiwu}, or via application of a bounded function to a random process,  e.g. \cite{bobryk,dimentberg}. We recall two examples which will be used in the applications, further details  are available in \cite{dOnofrioBoundedNoiseBook}.

\begin{figure}[t]
\centerline{
 	  \includegraphics[width=0.3\textwidth]{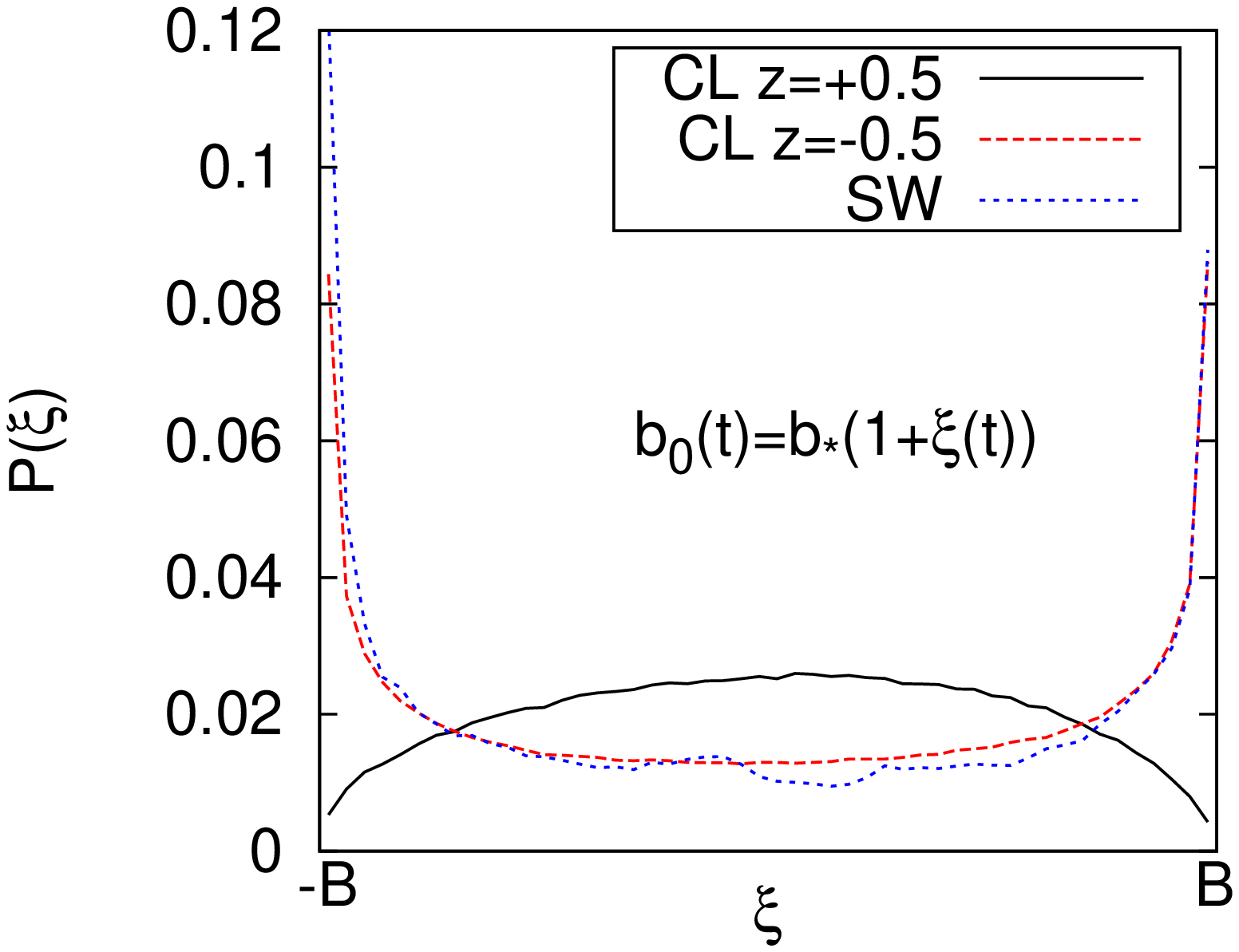}
}
	\caption{Stationary distributions of the Cai-Lin and the sine-Wiener bounded noises suggest different types of perturbations, according to the parameters.}
	\label{fig:noise}
\end{figure}

\paragraph{\bf Cai-Lin bounded noise.} Consider the folowing {\em Langevin equation}
\begin{equation}\label{Cai}
\dot{\xi} (t)= -\frac{1}{\tau}\xi(t) + \sqrt{\frac{B^2 -\xi^2}{\tau(1+z)}}\eta(t),
\end{equation}
where  $\eta(t)$ is a Gaussian white noise. As it is easy to verify, if $\xi(0) \in [-B,+B]$ then  $ \xi(t) \in [-B,+B]$. Moreover, it  has zero mean and the same stationary autocorrelation of the Ornstein-Uhlenbeck process, and its steady-state probability is 
$$ \Probab_{st}(\xi) = C (1-\xi^2)^z $$
 density exhibits different shapes according to $z$.  For $z>0$ it is unimodal and centered in $0$, while for $-1<z<0$ it is bimodal with two vertical asymptotes at $\xi\rightarrow \pm B$ (i.e., it is ``horned"). 
\\
\paragraph{\bf Sine-Wiener bounded noise.} This is obtained by applying a bounded function to a  Wiener process $W(t)$ yielding
\begin{equation}
\xi(t)= B \sin\left( \sqrt{ \frac{2}{\tau}}W(t) \right)\, .
\end{equation}
The noise steady-state distribution is horned and equal to the one of the Cai-Lin case for $z= - 0.5$.\\
In figure \ref{fig:noise} some typical steady-state distributions from Cai-Lin and sine-Wiener approaches are depicted.
\\
\paragraph{\bf Autocorrelation.}
Autocorrelation, also known as serial correlation, is the cross-correlation of a signal (a temporal series) with itself. Informally, it is the similarity between observations as a function of the time-lag between them. 
Let suppose the noise has at time $\hat{t}$ a value $\xi(\hat{t})$. Autocorrelation characteristic time $\tau$ measures the time window in which the noise has a tendency to ``remember'' its past history, so up to $\hat{t}+\tau$ the value of $\xi(t)$ is somewhat similar to $\xi(\hat{t})$. When the noise is totally uncorrelated, like is the case of Gaussian white noise, the value $\xi(\hat{t}+dt)$ is totally independent from $\xi(\hat{t})$ for any $dt$ (in other words one can say that $\tau\rightarrow 0$).\\    

\subsection{Analytical results on the model D}

Concerning the model D, we may give some analytical results of interest. Here we assume that the parameters of the FC model are such that the system is multistable. We define the utility functions $F(U)$ (with $U\ge b_l$), $G(U)$ (with $U\in [b_l,b_r]$) and $H(u)$ (with $U\le b_r$) which compute, respectively, the smallest, the intermediate and the largest of the three real solutions of the equation:
$$ s n \frac{c_0 + c_2 y^2}{U + c_0 + c_2 y^2}  - d y =0.$$ 
First suppose that $b_\ast(1-B)\ge b_l$. Thus, if $y(0) \in [0,G(b_\ast(1-B))]$ then for large times $y(t) \in [ F(b_\ast(1+B) ), F(b_\ast(1-B) )] $. This results follows from the following differential inequalities:
\begin{equation}\label{ineq}
s n \frac{c_0 + c_2 y^2}{b_\ast (1+B) + c_0 + c_2 y^2}  - d y \le y^{\prime} \le s n \frac{c_0 + c_2 y^2}{b_\ast (1-B) + c_0 + c_2 y^2}  - d y
\end{equation}
Indeed, the above inequalities imply that $y_a(t)<y(t)<y_b(t)$ where $y_a(t)$ solve the following ODEs:
$$y_a^{\prime}=s n \frac{c_0 + c_2 y_a^2}{b_\ast (1+B) + c_0 + c_2 y_a^2}  - d y_a, \; y_a(0)=y(0) $$
$$y_b^{\prime}=s n \frac{c_0 + c_2 y_b^2}{b_\ast (1-B) + c_0 + c_2 y_b^2}  - d y_b, \; y_b(0)=y(0). $$
As it is easy to verify, these ODEs are such that for large times $y_a\ge F(b_\ast(1+B) )$ and $y_b \le F(b_\ast(1-B) )] $. Similarly, if $b_\ast(1+B)\ge b_r$ and $y(0) \ge G(b_\ast(1+B))$ then for large times it must be $y(t) \in ( H(b_\ast(1-B)),H(b_\ast(1+B)) )$. Moreover, if 
$$ b_\ast(1-B) <b_l <b_\ast<b_\ast(1+B)<b_r $$ 
and $y(0)<F(b_l)$ it apparently follows the quite neutral result that for large times it must be $y(t) \in [H(b_\ast(1-B)),F(b_\ast(1+B))]$. However, if in a time instant $\widehat{t}$ it is $y(\widehat{t})\ge G(b_\ast(1+B))$ then (based on the above inequalities) it must be that for $t>\widehat{t}$ it is $y(t) \in  [H(b_\ast(1-B)),G(b_\ast(1+B))]$. Due to the random nature of the perturbations, the existence of such a $\widehat{t}$ is very likely, if not sure. Indicating with 
$$ \Probab_L(Y,t) =Prob( y(t) \in (Y,Y+dY) | y(0) \in [0,F(b_l)]   ) $$
and with $ \Probab_L^{st} (Y)= \Probab_L(Y,+\infty) $ the above results suggest that in the case $y(0) \in [0,F(b_l)]$: 
\begin{itemize}
\item if $b_\ast \in (b_l,b_r)$ and $b_\ast(1-B)>b_l$ then $\Probab_L^{st} (Y)$ it is null outside $[ F(b_\ast(1+B) ), F(b_\ast(1-B) )]$. 
\item if $b_\ast \in (b_l,b_r)$, $b_\ast(1-B)<b_l$, and $b_\ast(1+B)>b_r$ then $\Probab_L^{st} (Y)$ it is null outside $[ H(b_\ast(1+B) ), H(b_\ast(1-B) )]$
\end{itemize}
The second result suggests that defining the ``order parameter'' $\langle y\rangle$, and considering its variation with $B\ge 0$, it may undergo a fist order transition at $ b_\ast(1-B) = b_l$, because there $\langle y\rangle(B)$ is discontinuous. A similar result can be obtained with reference to the upper branch of the bifurcation diagram. Indeed, defining 
$$ P_H(Y,t) =Prob( y(t) \in (Y,Y+dY) | y(0) \ge F(b_r)   ) $$
and $ P_H^{st} (Y)= P_H(Y,+\infty) $  it follows that in the case $y(0) \ge F(b_r)$
\begin{itemize}
\item if $b_\ast \in (b_l,b_r)$ and $b_\ast(1+B)>b_r$ then $P_H^{st} (Y)$ it is null outside $[ H(b_\ast(1+B) ), H(b_\ast(1-B) )]$. 
\item if $b_\ast \in (b_l,b_r)$, $b_\ast(1+B)>b_r$, and $b_\ast(1-B)>b_l$ then $P_H^{st} (Y)$ it is null outside $[ F(b_\ast(1+B) ), F(b_\ast(1-B) )]$
\end{itemize}
With respect to the unconditioned probability density $P(Y,t)$ the above results show that its asymptotic behavior (in the
functional space of the probability densities, which is a subset of $\cal L^1$) strongly depends on its initial conditions.
Thus, in theory, in numerical simulations with $y(0) \in [0,F(b_l)]$ we would expect to observe such a first order
transition, i.e. that it exists a $B_c$ (with $b_\ast(1-B_c)=b_l$)  such that for $0\le B<B_c$ the stationary density
$\Probab_L^{st} (Y)$ is unimodal and located at low values of $y$, whereas for $B_c < B$ the density jumps at larger vales of $y$. Moreover for values $B>B_d$, where $b:m(1+B_d)=b_r$, we expect that $ \Probab_L^{st} (Y)$ gets bimodal and $\langle y\rangle (B)$ decreases. In other words, at $B_C$ there should be a fist-order transition, whereas at $B=B_d$ a noise-induced transition from unimodality to bimodality (and a smooth decrease of $\langle y \rangle(B)$). Two similar transitions should, thus, also be expected in the case $y(0) > F(b_r)$.

Finally, recently Verd et al. \cite{verd2014} introduced in the framework of Systems Biology models affected by \textit{deteterministic} time-varying perturbations the concept of time-varying Waddington's potential. We believe that this potential might be extremely useful also for stochastically perturbed systems as Model D. Namely, to our model:
$$
y^{\prime}= s \frac{n (c_0+c_2 y^{2})}{b_0(t)+c_0+c_2 y^{2}}  - d y,
$$
it is associated the following time-varying potential:
$$
W(y,t)= \frac{d}{2} y^2 - s n \left( y - \frac{b_0(t)}{\sqrt{b_0(t)+c_0}}\frac{1}{\sqrt{c_2}}Arctan\left( y \sqrt{\frac{c_2}{b_0(t)+c_0} }\right)  \right)
$$
The shape of the potential $W(x,t)$, and the number and basin of attraction of its 'potential holes', stochastically change in the time. Thus, for example, in case of small to moderate fluctuations of $b_0(t)$ the corresponding irreversible transition 'low to large' ('large to low') values of $y$ can heuristically be read as the irreversible 'capture' by large (low) value attractors of a trajectory initially confined in a potential hole centered at low values of $y$.


\begin{thebibliography}{10}

\bibitem{Ko}
M.~Ko, Journal of Theoretical Biology \textbf{53}, 181 (1991)

\bibitem{KeplerElston}
T.B. Kepler, T.C. Elston, Biophysical Journal \textbf{81}(6), 3116  (2001)

\bibitem{Lip2005}
P.~Paszek, T.~Lipniacki, A.R. Brasierc, B.~Tian, D.E. Nowakc, M.~Kimmel,
  Journal of Theoretical Biology \textbf{253}, 422 (2005)

\bibitem{Kaern}
M.K. M, T.~Elston, W.~Blake, J.~Collins, Nature Reviews Genetics \textbf{6},
  451 (2005)

\bibitem{Lip06}
T.~Lipniacki, P.~Paszek, A.~Marciniak-Czochra, A.R. Brasiere, M.~Kimmel,
  Journal of Theoretical Biology \textbf{238}, 348 (2006)

\bibitem{Lip13}
J.~Jaruszewicz, P.J. Zuk, T.~Lipniacki, Journal of Theoretical Biology
  \textbf{317}(0), 140 (2013)

\bibitem{LipP53}
K.~Puszynski, B.~Hat, T.~Lipniacki, Journal of Theoretical Biology
  \textbf{254}, 452 (2008)

\bibitem{ZH1}
T.S. Gardner, C.R. Cantor, J.J. Collins, Nature \textbf{403}(6767), 339 (2000)

\bibitem{BK}
N.I. Markevich, J.B. Hoek, B.N. Kholodenko, The Journal of cell biology
  \textbf{164}(3), 353 (2004)

\bibitem{Iannaccone}
L.~Wang, B.L. Walker, S.~Iannaccone, D.~Bhatt, P.J. Kennedy, T.T. William,
  Proceedings of the National Academy of Sciences \textbf{106}(16), 6638 (2009)

\bibitem{Zhdanov2}
V.P. Zhdanov, Chaos, Solitons \& Fractals \textbf{45}(5), 577 (2012)

\bibitem{Xiong}
W.~Xiong, J.E. Ferrell, Nature \textbf{426}(6965), 460 (2003)

\bibitem{ZH4}
H.H. Chang, P.Y. Oh, D.E. Ingber, S.~Huang, BMC cell biology \textbf{7}(1), 11
  (2006)

\bibitem{ZH3}
O.~Cinquin, J.~Demongeot, Journal of theoretical biology \textbf{233}(3), 391
  (2005)

\bibitem{verd2014}
B.~Verd, A.~Crombach, J.~Jaeger, BMC Systems Biology \textbf{8}(1), 43 (2014)

\bibitem{GlassKauffman}
L.~Glass, S.A. Kauffman, Journal of Theoretical Biology \textbf{39}(1), 103
  (1973)

\bibitem{Griffith}
D.A. Griffith, Journal of Statistical Planning and Inference \textbf{140}(11),
  2980 (2010)

\bibitem{Simon}
Z.~Simon, Journal of Theoretical Biology \textbf{8}(2), 258  (1965)

\bibitem{Thomas}
R.~Thomas, Journal of Theoretical Biology \textbf{73}(4), 631  (1978)

\bibitem{ThomasDAri}
R.~Thomas, R.~D'Ari, \emph{Biological Feedback} (Chapman \& Hall/CRC
  Mathematical \& Computational Biology, 1990)

\bibitem{IglesiasIngalls}
P.A. Iglesias, B.P. Ingalls (eds.), \emph{Control theory and systems biology}
  (MIT Press, 2010)

\bibitem{NoiPlos}
A.~Graudenzi, G.~Caravagna, G.~De~Matteis, M.~Antoniotti, PLoS ONE
  \textbf{9}(5), e97272 (2014)

\bibitem{dOnofrioBoundedNoiseBook}
A.~d'Onofrio, \emph{Bounded Noises in Physics, Biology, and Engineering}.
\newblock Modeling and Simulation in Science, Engineering and Technology
  (Springer New York, 2013)

\bibitem{detw}
P.B. Detwiler, S.~Ramanathan, A.~Sengupta, B.I. Shraiman, Biophysical Journal
  \textbf{79}(6), 2801 (2000)

\bibitem{RaoWolfArkin}
C.V. Rao, D.M. Wolf, A.P. Arkin, Nature \textbf{420}(6912), 231 (2002)

\bibitem{Thattai}
M.~Thattai, A.~van Oudenaarden, Biophysical journal \textbf{82}(6), 2943 (2002)

\bibitem{Becskei1}
A.~Becskei, L.~Serrano, Nature \textbf{405}(6786), 590 (2000)

\bibitem{hl}
W.~Horsthemke, R.~Lefever, \emph{Noise-Induced Transitions: Theory and
  Applications in Physics, Chemistry, and Biology (Springer Series in
  Synergetics)} (Springer, 2006)

\bibitem{hasty}
J.~Hasty, J.~Pradines, M.~Dolnik, J.J. Collins, Proceedings of the National
  Academy of Sciences \textbf{97}(5), 2075 (2000)

\bibitem{Arkin}
C.V. Rao, D.M. Wolf, A.P. Arkin, Nature \textbf{420}(6912), 231 (2002)

\bibitem{siggia}
M.B. Elowitz, A.J. Levine, E.D. Siggia, P.S. Swain, Science \textbf{297}(5584),
  1183 (2002)

\bibitem{Becskei2}
A.~Becskei, B.B. Kaufmann, A.~van Oudenaarden, Nature genetics \textbf{37}(9),
  937 (2005)

\bibitem{cfx}
L.~Cai, N.~Friedman, X.S. Xie, Nature \textbf{440}(7082), 358 (2006)

\bibitem{G77}
D.T. Gillespie, The journal of physical chemistry \textbf{81}(25), 2340 (1977)

\bibitem{ThattaiPNAS}
M.~Thattai, A.~Van~Oudenaarden, Proceedings of the National Academy of Sciences
  \textbf{98}(15), 8614 (2001)

\bibitem{To}
T.~Tze-Leung, N.~Mahesci, Science \textbf{327}, 1142 (2010)

\bibitem{eldar}
A.~Eldar, M.B. Elowitz, Nature \textbf{467}(7312), 167 (2010)

\bibitem{losick}
R.~Losick, C.~Desplan, science \textbf{320}(5872), 65 (2008)

\bibitem{bucetaplosone}
M.~Weber, J.~Buceta, PLoS ONE \textbf{8}(9), e73487 (2013)

\bibitem{GarciaOjalvo2012}
J.~Garcia-Ojalvo, A.M. Arias, Current Opinion in Genetics and Development
  \textbf{22}(6), 619  (2012).
\newblock Genetics of system biology

\bibitem{Tsimring}
L.S. Tsimring, Reports on Progress in Physics \textbf{77}(2), 026601 (2014)

\bibitem{msb08}
V.~Shahrezaei, J.F. Ollivier, P.S. Swain, Molecular Systems Biology
  \textbf{4}(1), n/a (2008)

\bibitem{GiulioDon12PONE12}
G.~Caravagna, G.~Mauri, A.~d'Onofrio, PLoS ONE \textbf{8}(2), e51174 (2013)

\bibitem{NoiNatComp}
S.~de~Franciscis, G.~Caravagna, A.~d'Onofrio, Natural Computing \textbf{13}(3),
  297 (2014)

\bibitem{dongan}
A.~d'Onofrio, A.~Gandolfi, Phys. Rev. E \textbf{82}, 061901 (2010)

\bibitem{jsgriffith}
J.S. Griffith, Journal of Theoretical Biology \textbf{20}(20), 209 (1968)

\bibitem{sbb}
P.~Smolen, D.A. Baxter, J.H. Byrne, American Journal of Physiology-Cell
  Physiology \textbf{274}(2), C531 (1998)

\bibitem{sbb99}
P.~Smolen, D.A. Baxter, J.H. Byrne, American Journal of Physiology - Cell
  Physiology \textbf{277}(4), C777 (1999)

\bibitem{frigolaplosone}
D.~Frigola, L.~Casanellas, J.M. Sancho, M.~Ibañes, PLoS ONE \textbf{7}(2),
  e31407 (2012)

\bibitem{PreSBB}
Q.~Liu, Y.~Jia, Physical Review E \textbf{70}, 041907 (2004)

\bibitem{PhysaSBB}
X.M. Liu, H.Z.X. ans Liang-Gang~Liu, Z.B. Li, Physica A \textbf{388}, 392
  (2009)

\bibitem{AssafPRL}
M.~Assaf, E.~Roberts, Z.~Luthey-Schulten, N.~Goldenfeld, Phys. Rev. Lett.
  \textbf{111}, 058102 (2013)

\bibitem{Stranger}
B.E. Stranger, M.S. Forrest, M.~Dunning, C.E. Ingle, C.~Beazley, N.~Thorne,
  R.~Redon, C.P. Bird, A.~de~Grassi, C.~Lee, C.~Tyler-Smith, N.~Carter, S.W.
  Scherer, S.~Tavar{\'e}, P.~Deloukas, M.E. Hurles, E.T. Dermitzakis, Science
  \textbf{315}(5813), 848 (2007)

\bibitem{Cappuzzo}
F.~Cappuzzo, F.R. Hirsch, E.~Rossi, S.~Bartolini, G.L. Ceresoli, L.~Bemis,
  J.~Haney, S.~Witta, K.~Danenberg, I.~Domenichini, V.~Ludovini, E.~Magrini,
  V.~Gregorc, C.~Doglioni, A.~Sidoni, M.~Tonato, W.A. Franklin, L.~Crino, P.A.
  Bunn, M.~Varella-Garcia, Journal of the National Cancer Institute
  \textbf{97}(9), 643 (2005)

\bibitem{Davis}
M.H. Davis, Journal of the Royal Statistical Society. Series B (Methodological)
  pp. 353--388 (1984)

\bibitem{G76}
D.T. Gillespie, Journal of computational physics \textbf{22}(4), 403 (1976)

\bibitem{wioII}
H.S. Wio, R.~Toral, Physica D: Nonlinear Phenomena \textbf{193}(1-4), 161
  (2004)

\bibitem{caiwu}
G.~Cai, C.~Wu, Probabilistic Engineering Mechanics \textbf{19}(3), 197  (2004).
\newblock Fifth International Conference on Stochastic Structural Dynamics

\bibitem{bobryk}
R.V. Bobryk, A.~Chrzeszczyk, Physica A: Statistical Mechanics and its
  Applications \textbf{358}(2-4), 263 (2005)

\bibitem{dimentberg}
M.~Dimentberg, \emph{Statistical dynamics of nonlinear and time-varying
  systems} (Research Studies Press, 1988)

\bibitem{RosenfeldScience}
N.~Rosenfeld, J.W. Young, U.~Alon, P.S. Swain, M.B. Elowitz, Science
  \textbf{307}(5717), 1962 (2005)

\bibitem{peliti2011statistical}
L.~Peliti, \emph{Statistical Mechanics in a Nutshell}.
\newblock In a Nutshell (Princeton University Press, 2011)

\bibitem{SoleBook}
R.V. Sol\'{e}, \emph{Phase Transitions} (Princeton University Press, 2011)

\bibitem{Alberts}
B.~Alberts, A.~Johnson, J.~Lewis, M.~Raff, K.~Roberts, P.~Walter,
  \emph{Molecular Biology of the Cell}, 5th edn. (Garland Science, 2009)

\bibitem{mcgraw}
M.H.P. Company, \emph{McGraw-Hill Concise Encyclopedia of Science and
  Technology} (McGraw-Hill, 2009)

\bibitem{Blake2006}
W.J. Blake, G.~Balazsi, M.A. Kohanski, F.J. Isaacs, K.F. Murphy, Y.~Kuang, C.R.
  Cantor, D.R. Walt, J.J. Collins, Molecular Cell \textbf{24}(6), 853  (2006)

\bibitem{Suter2011}
D.M. Suter, N.~Molina, D.~Gatfield, K.~Schneider, U.~Schibler, F.~Naef, Science
  \textbf{332}(6028), 472 (2011)

\bibitem{gutierrez2012}
P.S. Gutierrez, D.~Monteoliva, L.~Diambra, PLoS ONE \textbf{7}(9), e44812
  (2012)

\bibitem{wiolindenberg}
H.S. Wio, K.~Lindenberg, Modern Challenges in Statistical Mechanics. AIP
  Conference Proceedings \textbf{658}, 1 (2003)

\bibitem{Gardiner}
C.W. Gardiner, et~al., \emph{Handbook of stochastic methods}, vol.~3 (Springer
  Berlin, 1985)

\bibitem{Hanggi}
P.~Jung, P.~H{\"a}nggi, Phys. Rev. A \textbf{35}, 4464 (1987)

\bibitem{Hanggi2}
P.~H{\"a}nggi, \emph{Colored Noise in Continuous Dynamical Systems: A
  Functional Calculus Approach}, vol. 1609 (Cambridge University Press, 1989)

\bibitem{G80}
D.T. Gillespie, The Journal of Chemical Physics \textbf{72}(10), 5363 (1980)

\bibitem{Gillespie00}
D.T. Gillespie, The Journal of Chemical Physics \textbf{113}(1), 297 (2000)

\bibitem{You}
M.~Hallen, B.~Li, Y.~Tanouchi, C.~Tan, M.~West, L.~You, PLoS Comp. Biol.
  \textbf{7}(10), e1002209 (2011)

\bibitem{hilfpaul}
A.~Hilfinger, J.~Paulsson, Proceedings of the National Academy of Sciences
  \textbf{108}(29), 12167 (2011)

\end{thebibliography}
\end{document}